\documentclass[structabstract]{aa}      %%%<<<<<<<<<<<<<<<<<<
\usepackage{natbib} %%%<<<<<<<<<<<<<<<<<<
\bibpunct{(}{)}{;}{a}{}{,} % to follow the A&A style
\usepackage{graphicx}
\usepackage{txfonts}
\begin{document}
   \title{The mid-IR extinction in molecular clouds.}

   \subtitle{Case study of B 335}

  \author{S. Olofsson \inst{1}
          \and
          G. Olofsson \inst{1}
          \fnmsep \thanks{Based on observations collected at the European Southern Observatory,
          Chile (ESO programs 073.C-0436(A), 077.C-0524(A))
          and on observations made with the Nordic Optical Telescope (NOT),
          operated on the island of La Palma jointly by Denmark, Finland, Iceland,
          Norway, and Sweden, in the Spanish Observatorio del Roque de los Muchachos
         of the Instituto de Astrofisica de Canarias.}
         }
   \institute{Stockholm Observatory, Stockholm University, Astronomy Department
              AlbaNova Research Center, SE-106 91 Stockholm\
              \email{sven@astro.su.se}
    }
   \date{Received        ;accepted       }

% \abstract{}{}{}{}{}
% 5 {} token are mandatory
\abstract
   {}
   {The purpose of the present investigation is to probe the dust properties
    inside a molecular cloud, in particular how particle growth
    and the presence of ice coatings may change the overall shape of the extinction curve.}
   {Field stars behind a molecular cloud can be used to probe the cloud extinction
    for both the reddening and the absorption features.
    By combining multi-colour photometry and IR spectroscopy
    the spectral class of the star can be determined as can the extinction curve,
    including the vibrational bands of ices and silicates. }
   {Based on observations of field stars           %%a M giant star and an A star
    behind the dark globule B\,335,
    we determine the reddening curve from 0.35 to 24\,$\mu$m.
    The water ice band at 3.1\,$\mu$m is weaker ($\tau$(3.1) = 0.4)
    than expected from the cloud extinction (A$_{V} \approx $  10 for the sightline to the most obscured star).
    On the other hand, the CO ice band at 4.7\,$\mu$m is strong ($\tau$(4.67) = 0.7)
    and indicates that the mass column density of frozen CO is about the same as that of water ice.
    We fit the observations to model calculations and find that the thin ice coatings
    on the silicate and carbon grains (assumed to be spherical) \textit{lower}
    the optical  extinction by a few percent.
    We show that the reddening curves for the two background stars,
    for which the silicate band has been measured, can be accurately modelled from the UV to $24\ \mu$m.
    These models only include  graphite and silicate grains (plus thin ice mantles for the most obscured star),
    so there is no need for any additional major grain component
    to explain the slow decline of the reddening curve beyond the K band.
    As expected, the dust model for the dense part of the cloud has more  large grains than for the outer regions.
    We propose that the well established shallow reddening curve beyond the K band has two different explanations: larger graphite grains in dense regions and relatively small grains in the diffuse ISM,
    giving rise to substantially less extinction beyond the K band than previously thought.}
   { For the sightline towards the most obscured star,
   we derive the relation A$_{Ks}$ = 0.97$\cdot$E(J$-$K$_{Ks}$),
   and assuming that all silicon is bound in silicates,
   N(2\,H$_2$+H) $\approx \  1.5\cdot 10^{21} \cdot A_{V}\ \approx \  9\cdot 10^{21} \cdot A_{Ks}$.
   For the rim of the cloud we get  A$_{Ks}$ = 0.51 $\cdot$E(J$-$K$_s$),
   which is close to recent determinations for the diffuse ISM.
   The corresponding gas column density is
   N(2\,H$_2$+H) $\approx \  2.3\cdot 10^{21} \cdot A_{V}\ \approx \  3\cdot 10^{22} \cdot A_{Ks}$.}

   \maketitle
\section{Introduction}

\textit{Interstellar reddening} is caused by small dust particles,
but the detailed optical and chemical properties of these particles
remain largely unknown, in particular for molecular clouds.
With the exception of the  220 nm extinction peak (probably caused by small graphite grains),
%and the weak DIBs (diffuse interstellar bands) of unknown origin,
the UV, optical, and near-IR spectral regions offer little information about the dust particles,
except that they must be small compared to the wavelength (even in the UV).
In the mid-IR, however, strong absorption bands show up,
the most universal of these being the broad band centred at 9.7\,$\mu$m due to silicates.
Another widespread component in the \textit{diffuse} ISM (interstellar medium),
PAHs (polycyclic aromatic hydrocarbons) are seen in the mid-IR as emission (due to luminescence).
In addition, the far-IR thermal emission from the interstellar dust grains
gives further constraints on their size distribution, and to some extent, their composition.\\
\\
Combining all this information, \cite{2003ARA&A..41..241D} has constructed a dust model        %% Draine 2003
that describes the observed properties for the diffuse ISM.
It basically consists of silicate and carbon grains. For molecular clouds,
the lower temperature, the higher density and
a diluted UV radiation field allow ices to form and/or condense on the grains.
In addition, the grains coagulate and the size distribution is changed towards larger grains.
These processes alter the optical properties of the grains.
The most obvious difference compared to the diffuse ISM is the absorption features from ices.
The ISO instruments provided extensive spectroscopic evidence of various ice absorption features,
including H$_2$O, CO, CO$_2$, CH$_4$, CH$_3$OH and NH$_3$ \citep{2004ApJS..151...35G}.
In most cases, these absorption features have been observed
towards luminous, deeply embedded protostars,
and it is hard to separate the circumstellar matter
(influenced by dynamical and radiation processes
involved in the formation of the star) from the unprocessed cloud component.
In addition, the intrinsic SED (spectral energy distribution) of an embedded young star
is not in general well known, which means that it is difficult to accurately
separate interstellar reddening from intrinsic IR excess.
For these reasons, field stars behind a molecular cloud are better
probes in investigations of the (more or less) undisturbed cloud.
The problem is to find background stars that are heavily obscured to probe
the dense regions and still bright enough for IR spectroscopy.
In spite of these difficulties, there are some investigations of ice features using background stars:
Elias 16 by \cite{2004ApJS..151...35G} and
Spitzer ice-data from three sources behind Taurus and CK2 in Serpens by \cite{2005ApJ...635L.145K}. More recently, \cite{2011ApJ...729...92B} has combined ground-based and
Spitzer spectroscopy data to analyse the ice and silicate features
in quiescent clouds by observing background field stars. \\

Apart from the absorption bands, the continuous extinction
in the mid-IR for molecular clouds has been the subject of several recent investigations
\citep[e.g.][]{2007ApJ...663.1069F, 2007ApJ...664..357R, 2009ApJ...690..496C, 2011ApJ...729...92B}.
However, it is important to note that what is actually determined is the {\it reddening curve},
and to establish the {\it extinction curve}, an assumption is made on the extinction ratio for two wavelengths, e.g. A$_H$/A$_K$ = 1.55, which goes back to the frequently cited paper by \cite{1985ApJ...288..618R}. Their paper combined earlier investigations with mid-IR observations of distant field stars of unknown intrinsic colours. Obviously the data available for their analysis was very limited compared to those in recent investigations, based on large surveys both from ground and space.
The reddening curve is relatively straight forward to determine, either by statistical methods (slopes in colour/colour diagrams) or by classifying selected background stars and compare the observed colour to the intrinsic. By observing stars all the way to Spitzer/MIPS 24 $\mu$m one would expect that the extinction is so close to zero at this wavelength, that it for all practical purposes does not matter if it is assumed to be exactly zero. However, the observations show that the reddening curve levels off in the mid-IR, which is not expected from current dust models, and for this reason the extrapolation to zero wave number remains uncertain.
\\
In our recent paper \citep[][hereafter called Paper\,II]{2010A&A...522A..84O}
we confirmed that the reddening curve for
molecular clouds in the wavelength range 0.35--2.2 $\mu$m can be characterized in a functional  form described by
\cite{1989ApJ...345..245C} (henceforth CCM) with only one parameter,  $R_{V}\,=\,A_{V}/E_{B-V}$. It was, however, clear that the CCM function did not adequately describe the reddening curve beyond the K band, and
in the present paper we extend the investigation into the mid-IR spectral region.
For our case study we have compiled our own and archive data
for a few stars behind the B\,335 globule including photometry and spectroscopy
from the optical region to 24\,$\mu$m.
Our main purpose is to characterize the mid-IR extinction.
In addition we explore whether both the reddening curve, from UV to 24\,$\mu$m, and
the absorption features can be explained in a single model. \\
\\
\begin{table*}
\begin{minipage}[t]{\columnwidth}
\caption{Observations and data.}
\label{obs}
\renewcommand{\footnoterule}{}  % to avoid a line before footnotes
\begin{tabular}{llllll}
\hline
%\textbf{\textbf{band}}             & U602  & Bb605  & g772  & V606  & r773  & I610  & 2massJ & 2massH & 2massKs&$[3.6\mu]$&$[5.8\mu]$\\
%$\lambda\ \ \ [nm]$                & 354.0 & 413.2  & 508.9 & 542.1 & 673.3 & 798.1 & 1258.5 & 1649.7 & 2157.2 &  & \\
%$\Delta\lambda\ [nm]$              &  53.7 & 109.4  &  75.3 & 104.8 &  81.1 & 155.2 &  300.  &  300.  &  300.  & 800. & 1500.\\
\hline
\textbf{observation}&\textbf{band}& \textbf{observatory} & \textbf{wavelengths $\mu m$}& \textbf{observation date}& \textbf{reference}\\
\hline
imaging          & optical      & NTT-EMMI     & [0.354,\,0.413,\,0.509,\,0.673,\,0.798] & 2006-06-27--29 \\ % & 077.C-0524(A) \\
spectroscopy     & optical      & NTT-EMMI     & $0.5\,<\lambda<\,0.8$                   & 2006-06-27--29 \\ % & 077.C-0524(A) \\
imaging          & NIR          & 2MASS        & [1.258,\,1.649,\,2.157]                 &                  & \cite{2006AJ....131.1163S}\\  %% Skrutskie 2006
imaging          & L'           & NOT-SIRCA    & 3.79                                    & 2003-05-17       & unpublished \\
spectrometry     & H$_2$O\,ice  & VLT-ISAAC    & $2.55\,<\lambda<\,4.2$                  & 2004-04-12--13 \\    %& 073.C-0436A.1\\
spectrometry     & CO\,ice      & VLT-ISAAC    & $4.45\,<\lambda<\,5.1$                  & 2004-04-12--13 \\    %& 073.C-0436A.1\\
%%
%imaging          & NIR          & Spitzer-IRAC & $3.6,\ 4.5,\ 5.8,\ 8.0$                 & 2004-04-20       & AOR:4926208 IRAC-0000-CB199\\
%imaging          & NIR          & Spitzer-IRAC & $3.6,\ 4.5,\ 5.8,\ 8.0$                 & 2004-04-21       & AOR:4926464 IRAC-0000-CB199\\
%%%imaging          & NIR          & Spitzer-MIPS & 24.0                                    & 2004-10-16       & AOR:12022016 MIPSC-CB199(B335)\\
%%%
spectrometry     & silicate     & ISOCAM-CVF   & $5\,<\lambda<\,16.5$                    & 1997-04-25       & Rev-TDT-OSN: 526 003 74\\
spectrometry     & silicate     & Spitzer-IRS  & $5\,<\lambda<\,21 $                     & 2005-10-11       & AOR:10705664 VeluCores5\\
CO isotopologues & mm           & ESO-SEST     & $^{13}$CO and C$^{18}$O   1--0          & 1993 -- 1998     & Harjunp\"{a}\"{a} et al (2004)\\ %%\footnote{\cite{2004A&A...421.1087H}}\\               %%Harjunp\"{a}\"{a} et al, 2004\\
\hline
\end{tabular}
\end{minipage}
\end{table*}
%%ISOCAM/CVF
%%Summary Details
%%---------------
%%
%%Target Name       :B335 CVF
%%Inst/AOT          :CAM04
%%Position          :294.254122 7.569690
%%Dimensions        :f.o.v. 192"x192"
%%Wavelengths       :5.01-16.77µm
%%Rev-TDT-OSN       :526 003 74
%%Observer          :SCABRIT
%%Proposal          :CAMFLOWS
%%Start Time        :1997-04-25 05:20:07.245
%%End Time          :1997-04-25 06:48:07.245
%%On Target Time [s]:5280 sec
%%Quality           :Good

\section{Observations and reductions}
\begin{figure}
   \centering
   \includegraphics[width=9cm]{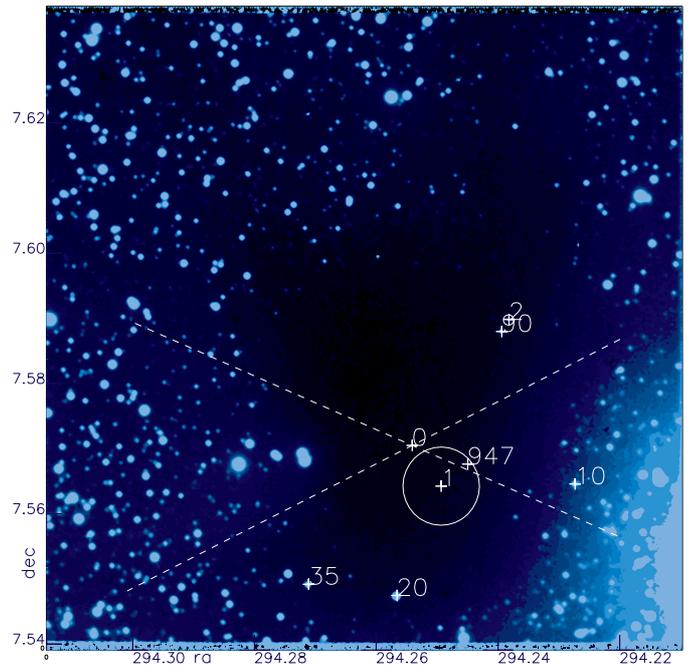}
      \caption{A U band image of the B335 globule. The centre YSO and its outflow region
      as well as the background stars are marked. The circle represents
      the position of the $^{13}$CO and the C$^{18}$O measurement
      closest to the background star \#947  in \cite{2004A&A...421.1087H}
      and the size of the circle represents the 45" beam width.
%%      The numbering of stars are the same as in \emph{paper II}
      }
      \label{B335}
   \end{figure}

The present investigation is based on observations of
a few background stars behind the B\,335 dark globule
(RA(2000)\,=\,[19 37 00], DEC(2000)\,=\,[7 34 10]) in both the optical and the near-IR.
%(RA(2000)\,=\,294.25, DEC(2000)\,=\,7.57) in both optical and NIR.
Their positions are marked in Fig \ref{B335}.
Table \ref{summary} shows the observations available in our investigation.
\\
\begin{table*}
\begin{minipage}[t]{\columnwidth}
\caption{Summary of the background stars and available measurements.}
\label{summary}
\renewcommand{\footnoterule}{}  % to avoid a line before footnotes
\begin{tabular}{lllllllll}
\hline\hline\\
 & \textbf{central star} &\multicolumn{6}{c}{\textbf{background stars}\footnote{numbers refer to star positions in Fig \ref{B335}.}}\\
%& $^{\circ}K$ & J2000 & J2000        & 0.35 & 0.41  & 0.51 & 0.67 & 0.80 & 1.26 &1.65 & 2.16 & 3.79 & 3.6 & 4.5 & 5.8 & 8 & 24& 5 - 22 \\
%$\lambda\ \ \ [$\mu m$]              & 354.0 & 413.2 & 508.9 & 542.1 & 673.3 & 798.1 & 1258.5 & 1649.7 & 2157.2 &  & \\
%$\Delta\lambda\ [nm]$                &  53.7 & 109.4  &  75.3 & 104.8 &  81.1 & 155.2 &  300.  &  300.  &  300.  & 800. & 1500.\\
\hline
star \#        &  YSO      & 2        & 10       & 20       & 84       & 90       & 112      & 947     \\
ra             &  19 37 00 & 19 36 57 & 19 36 54 & 19 37 01 & 19 36 56 & 19 36 57 & 19 36 57 & 19 36 58\\
dec            &  07 34 10 & 07 35 23 & 07 33 49 & 07 32 43 & 07 33 44 & 07 35 16 & 07 33 30 & 07 33 59\\
$T_{eff}$      &           & 3800     & 4000     & 9200     & 6000     & 9800     & 6400     & 3050    \\
\hline\hline
\textbf{photometry}&\textbf{central}&\multicolumn{6}{c}{\textbf{flux $[mJy]$}}\\
                   &\textbf{wavelength $[\mu m]$}\\
\hline
\textbf{NTT-EMMI}\\
U602      & 0.354  &         0.004    & 0.010      & 0.074   & 0.003   & 0.005    & 0.003   & .\\
Bb605     & 0.413  &         0.12     & 0.22       & 0.35    & 0.021   & 0.035    & 0.019   & .\\
g772      & 0.509  &         0.71     & 0.92       & 0.85    & 0.08    & 0.087    & 0.06    & .\\
r773      & 0.673  &         11.5     & 7.7        & 4.8     & 0.69    & 0.54     & 0.49    & 0.026 \\
I610      & 0.798  &         23       & 11         & 6.1     & 1.0     & 0.75     & 0.73    & 0.34 \\
\multicolumn{2}{l}{\textbf{2MASS}}\\
J         & 1.258  &         257      & 55         & 29      & 5.9     & 4.8      & 4.6     & 73\\
H         & 1.650  &         612      & 85         & 39      & 10      & 6.8      & 7.2     & 357\\
K$_{s}$   & 2.157  &         661      & 78         & 35      & 9.3     & 7.1      & 6.5     & 627\\
\textbf{SIRCA} & 3.79 &      .        & .          & .       & .       & .        & .       & 453\\
\multicolumn{2}{l}{\textbf{Spitzer}}\\                                                  %% IRAC_MIPSphot20100509.tab
IRAC1     & 3.550  &         .        & 38         & 15      & 4.3      & 3.6     & 3.0      & .  \\
IRAC2     & 4.493  &         220      & 22         & 9.8     & 2.8      & 2.5     & 2.0      & .  \\
IRAC3     & 5.731  &         169      & 16         & 6.7     & 1.5      & 1.9     & 1.3      & 243\\
IRAC4     & 7.872  &         93       & 8.9        & 3.5     & 0.8      & 1.2     & 0.7      & 149\\
MIPS      & 23.7   &         9.9      & 0.85       & 0.2:    & .        & 0.1:    & 0.1:     & 35\\
\hline\hline
\textbf{spectrometry} &\textbf{wavelength} & \multicolumn{6}{c}{\textbf{spectral resolution R}}\\
                      &\textbf{range $[\mu m]$}\\
\hline
NTT-EMMI    & $0.4<\lambda<0.8$   & +    & +       & +       & R$\simeq$400 &  +  & +        & +\\
VLT-ISAAC   & $2.55<\lambda<4.2$  & .    & .       & .       & .        & .       & .        & $R\simeq$600\\
VLT-ISAAC   & $4.45<\lambda <5.1$ & .    & .       & .       & .        & .       & .        & $R\simeq$3300\\
ISOCAM-CVF  & $5<\lambda <16.5$   & .    & .       & .       & .        & .       & .        & $100<R<240$\\
Spitzer-IRS & $\,7<\lambda <14$   & $60<R<120$ & . & . & .  & .  & .        & .\\
Spitzer-IRS & $14<\lambda <21$    & $60<R<120$ & . & . & .  & .  & .        & $60\ <R<120$\\
\hline\\
\end{tabular}
\end{minipage}
\end{table*}    %%%%%~/not9/lysTab/not99_B335matrixTeCOLARv1noptTeCOLARv1EMMIn2massnIRACS3NTeG60NRvG2_20090917wS947.tab
\begin{itemize}

\item Imaging and spectroscopy using the NTT/EMMI
at ESO, La Silla (077.C-0524(A) during four nights 2006-06-27--29.
These observations and reductions are described
in a previous article, \cite{2009A&A...498..455O} (Paper\,I).\\

\item Spectroscopy of the background star (\# 947)
in the water ice band at $2.7\,<\lambda<\,3.8\,\mu m$
and the CO-ice band at $4.0\,<\lambda<\,4.5\,\mu m$
using VLT-ISAAC at ESO, Paranal  during two nights, 2004-04-12--13.
A standard star HR7519 is observed before and after each observation of the program star.

The 2D-spectra are first corrected for bias and flat-field and
the 1D-spectra are extracted with the IRAF program pipeline 'APALL'.
The program and standard spectra are then divided by respective stellar atmospheric models.
These are then divided to correct for telluric atmospheric extinction.\\

\item Imaging at $3.79\,\mu m$ using SIRCA (Stockholm Observatory IR CAmera)
mounted on the  2.5 m Nordic  telescope on La Palma.\\
%%SIRCA is described in \cite{2000nott.work..151V}\\              %% van der Bliek 2000\\

\end{itemize}

These data have been combined with data from the ISO and the Spitzer archives
as summarized in Tables \ref{obs} and \ref{summary}.
%_____________________________________________________________
%   Bokglobubule B335 from
%%  excessMin.pro
%%  funFlag		= [0, 0, 1, 0]		;; [AlambdExc & excessM, WDdistribution, Ubackground, ice-analysis]
%%	bubbleFlag	= [1, 0, 0]		;; [gauss, 2*gauss, polynom]
%%  B335inU602bgCMYK2_20110921.ps                                figure 1
%-------------------------------------------------------------
%_____________________________________________________________
%_____________________________________________________________
%   S947 extinction from
%   is4S947opticalXt.pro --> is4S947opticalXt(1) psFlag = 1
%   C:\is4\ps\S947extrapolation_20100302.ps                figure 2
%   C:\is4\ps\S947extrapolation_20101006.ps
%   C:\is4\ps\S947extrapolation_20110124.ps
%-------------------------------------------------------------
   \begin{figure}
   \centering
   \includegraphics[width=8cm]{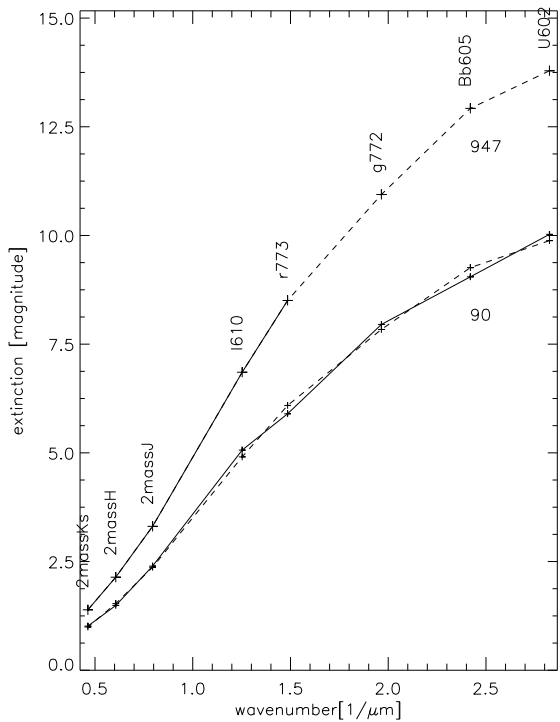}
      \caption{The optical and near-IR extinctions towards the background stars \#947 and \#90.
      The extinction for \#90 and \#947 are well represented by the fitted CCM-extinction with R$_V$\,=\,4.9 (dashed).
       }
      %For the two shortest wavelengths (where \#947 is not detected),
      %the extinction of \#947 is assumed to have the same shape as that of \#90.}
      \label{extrapol}
   \end{figure}

%_____________________________________________________________
%   compare #947 with Hauschildt3000
%   em6_componeDspecwHiRes2010.pro    idSpec =  11; iTeff = 2
%   S947andHau3000_20101011.ps
%   S947andHiRes3000_20101026.ps
%   S947andHiRes3000_20110121.ps                           figure 3
%-------------------------------------------------------------
   \begin{figure}
   \centering
   \includegraphics[width=8cm]{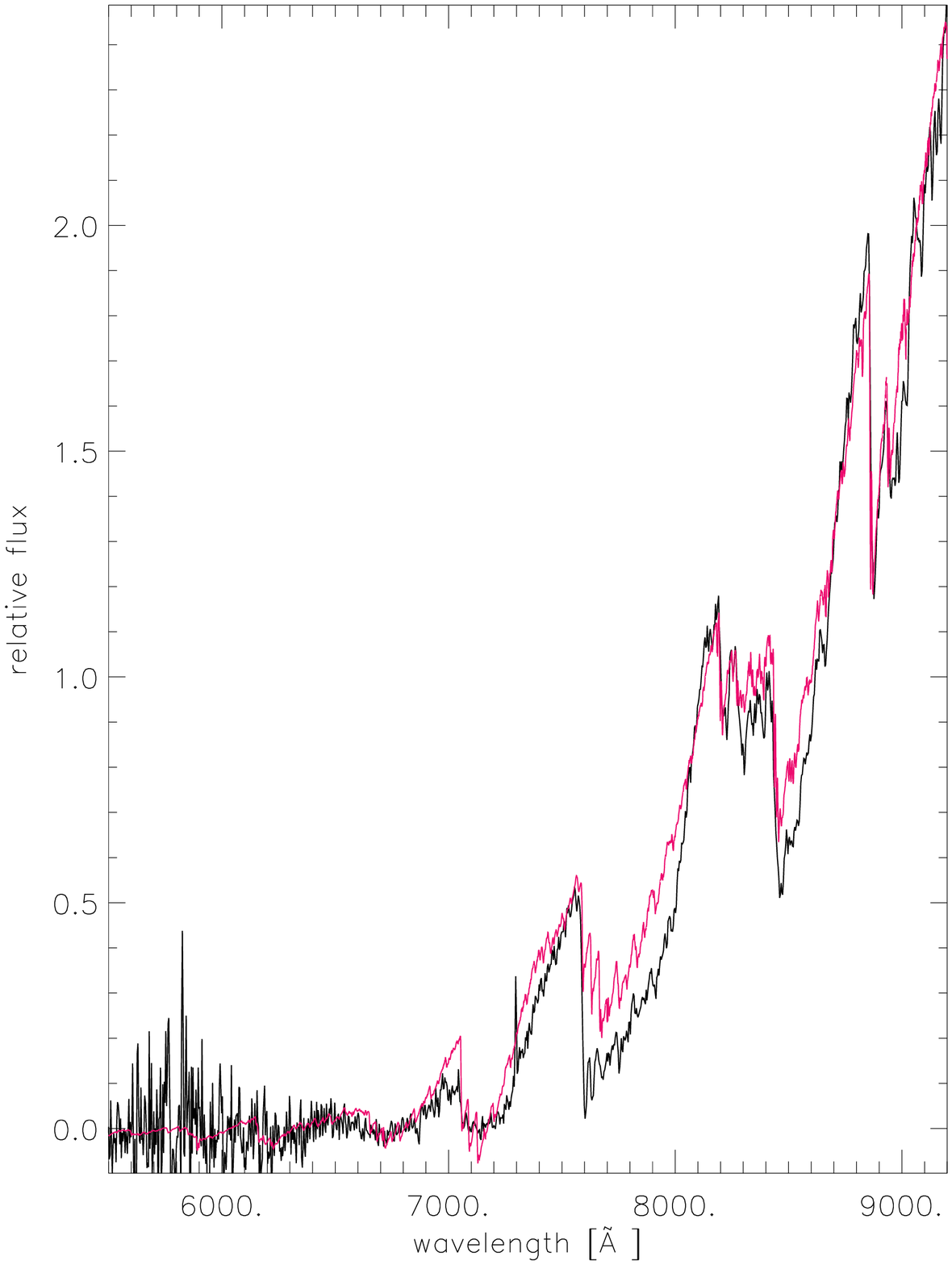}
   \caption{The optical spectrum of the most obscured star (\#947)
       is well represented by a stellar model \citep[from][]{1999ApJ...525..871H} with $T_{eff}$ 3050 K.
       The stellar model (\emph{red} curve) has been modified
       by a weighting function $\propto\ \lambda^{-1}$ to account for the interstellar reddening}
   \label{corr947}
   \end{figure}
%_____________________________________________________________
%
\section{Results}
\subsection{The intrinsic colours of the background stars}
In Paper\,II we confirmed that the molecular cloud reddening
in the optical and the near-IR spectral region
is well characterized by a functional form defined by CCM
with only one parameter, $R_{V}\,=\,A_{V}/E_{B-V}$.
This also allows us to use the multi-band photometry
to both determine the R$_V$ value and the spectral class for an obscured star as well as the reddening towards it.
For the two most obscured stars in the present study, \#947 and \#90, we get the same R$_{V}$ value,
$R_{V}\ =\ 4.9$ (Fig. \ref{extrapol}), and their spectral classes are M6III and A0V respectively.
The spectral class of the most obscured star in our sample, \#947, is of particular importance as it is the only
star for which we have water and CO ice data and the temperature determined by the multiband photometry
is confirmed both
by the optical spectrum (Fig. \ref{corr947}) and the L window spectrum (Fig. \ref{H2Oice}).
The spectral classes of the other stars are also confirmed by optical spectra.  %%, except \#NNNN (for which we lack spectroscopy).
This means that we can characterize the intrinsic SEDs (spectral energy distributions) of the stars,
and for that purpose we use model atmospheres from  \cite{1999ApJ...525..871H}.
\subsection{The H$_2$O and CO ice bands}
\begin{figure}                  %S947iceH2OTeffModel30_20110122.eps
    \centering
    \includegraphics[width=9cm]{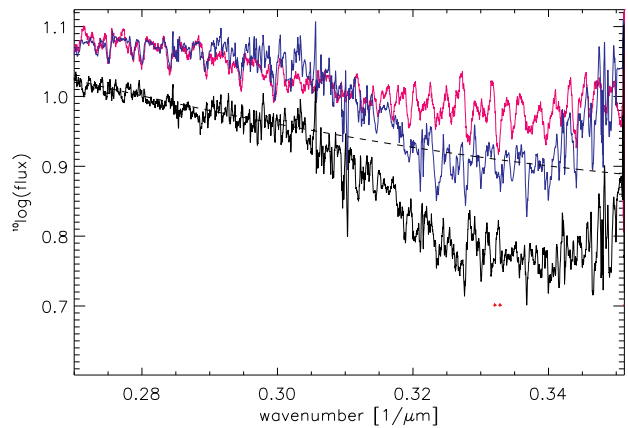}
       \caption{The ISAAC spectrum in the L band of star \#947 (\emph{blue} curve)
       and the stellar atmospheric model (\emph{red} curve) are divided to get the water ice band
        (\emph{black curve}).}
       \label{H2Oice}
    \end{figure}

\begin{figure}                  %S947iceCOTeffModel30_20110122.eps
    \centering
    \includegraphics[width=9cm]{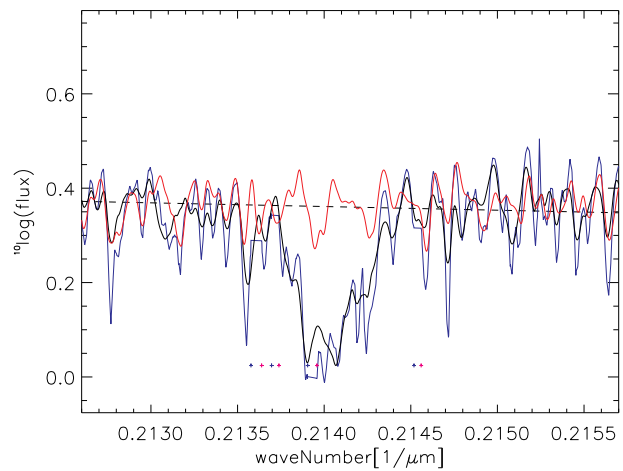}
       \caption{The ISAAC spectrum in the M band of star \#947 (\emph{blue} curve)
       and the corresponding stellar atmospheric model (\emph{red} curve) are divided to get the CO ice band
        (\emph{black curve}).}
       \label{COice}
    \end{figure}

By dividing the ISAAC spectra of star \#947 (corrected for telluric absorption and relative flux calibrated)
with a model atmosphere corresponding to the spectral class of the star (T$_{eff}$ = 3050 K and log\,g\,=\,0)
we determine the absorption of the H$_2$O and CO ice bands
(see Figs. \ref{H2Oice} and \ref{COice} respectively).
The peak position of the water ice band is at 3280\,cm$^{-1}$ (= 3.05\,$\mu$m)
with an FWHM of $350\ cm^{-1}$.
We note that there is no clear extension shortward of 3000\,cm$^{-1}$ (longward of 3.33\,$\mu$m).
Such an extension is often observed and explained as due to methanol ice \citep{2004ApJS..151...35G}.  %% Gibb, Whittet, Boogert & Tielens2004
The CO absorption peaks at 2140\,cm$^{-1}$ (= 4.67\,$\mu$m),
which differs from that of pure CO ice \citep[$\sim\ 2142\ cm^{-1}$,][]{1997ApJ...479..818E},      % Elsila & 1997 << flera referenser
but agrees with observations of other regions \citep{2003A&A...408..981P}.                       % Pontoppidan 2003
There is, however, no indication of wing components, although our S/N ratio
does not allow any firm statement on this point.\\
\\
%_____________________________________________________________
%   H$_{2}$Oice band spectra from is4spec2009.pro
%   SpecNumber  = 60
%   displayFlag = 1     wvnumOverLim    = [3000.,3510.]
%   wavenumFlag = 1     wvnumSampldx    = [3015., 3510.]
%   is4spec_2009.pro C:\is4\ps\S947H2OiceXtwvNum_20100928.eps
%                    C:\is4\ps\tauS947H2Oice_20101002.eps
%-------------------------------------------------------------
%    \begin{figure}
%    \centering
%    \includegraphics[width=9cm]{./bilder/tauS947H2Oice_20101002.eps}
%       \caption{Water ice band spectrum towards star \#947.}
%       \label{H$_{2}$Oicewvnum}
%    \end{figure}
%_____________________________________________________________
% H$_{2}$O ice from spectra is4H2Ospec2010
%   SpecNumber  = 60
%   displayFlag = [0,0,0,1,0]   wvnumOverLim    = [3000.,3510.]
%   wavenumFlag = 1             wvnumSampldx    = [3015., 3510.]
%   iCont       = [2,3,4,5,6,7,8,14]
%   modlSubtractdx[1,Ncont-1]	= HiResFlag ? -4.42
%   is4spec_2009.pro C:\is4\ps\S947iceH2OiceTeffModel30_20110121.eps    figure 4
%-------------------------------------------------------------
%_____________________________________________________________
%_____________________________________________________________
%   CO ice from spectra is4spec2009
%   SpecNumber  = 130
%   displayFlag = 1   wvnumOverLim    = [2134., 2145.]
%   wavenumFlag = 1   wvnumSampldx    = [2126., 2157.]
%   iXt       = [4,5,8,9,10,11,12]
%   is4spec_2009.pro C:\is4\ps\S947iceCOTeffModel30_20110122.eps    figure 5
%-------------------------------------------------------------
%_______________________________________________________
%   is4SpitzerExcess.pro
%   displayFlag = [1, 0, 0, 0]
%   SiFeature   = 0
%   in plot, /xlog                                              figure 6
%-------------------------------------------------------------
   \begin{figure}
   \centering
   \includegraphics[width=8cm]{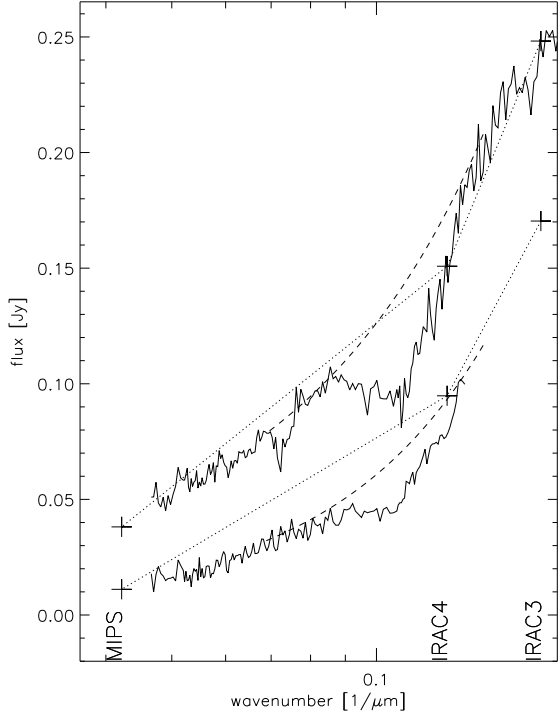}
      \caption{Spectra for sight lines towards stars \#2 (lower) and \#947 (upper curve)
      as observed by Spitzer-IRS (\#2: $7\ <\ \lambda \ < 14\ \mu m $)
      and ISOCAM-CVF (\#947: $5\ <\ \lambda \ < 16\ \mu m $) and
      photometry from Spitzer-IRAC and Spitzer-MIPS (+ signs).}
      \label{fluxes}
   \end{figure}
%%_____________________________________________________________
%_______________________________________________________
%   is4SpitzerExcess.pro
%   displayFlag = [1, 0, 0, 0]
%   SiFeature   = 1
%   in plot     ;;, /xlog
%%------------------------------------------------------------- figure 7
   \begin{figure}
   \centering
   \includegraphics[width=8cm]{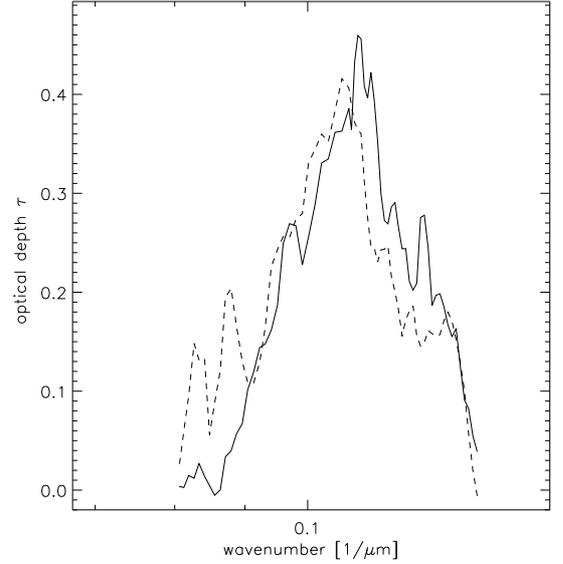}
      \caption{The optical depth of the silicate band $\tau_{silicate}$ towards star \#2 (\emph{dashed} line)
      and \#947 (\emph{full} line) as determined from Spitzer-IRS and ISOCAM-CVF observations, respectively.
      Although the S/N of the observations not allows for an interpretation of the details,
      we note that  for star \#947 the band peaks  at $\approx$ 0.109 $\mu$m$^{-1}$ corresponding to 9.2 $\mu$m.
      }
      \label{tauSi}
   \end{figure}
%_____________________________________________________________
%   extinction and extrapolation to zero wavenumber.
%   is4SpitzerExcess
%   displayFlag = [1,0,0,0]
%   ishow = 8           %%SpitzerExcess820110122.eps          figure 8
%-------------------------------------------------------------
   \begin{figure}
   \centering
   \includegraphics[width=8cm]{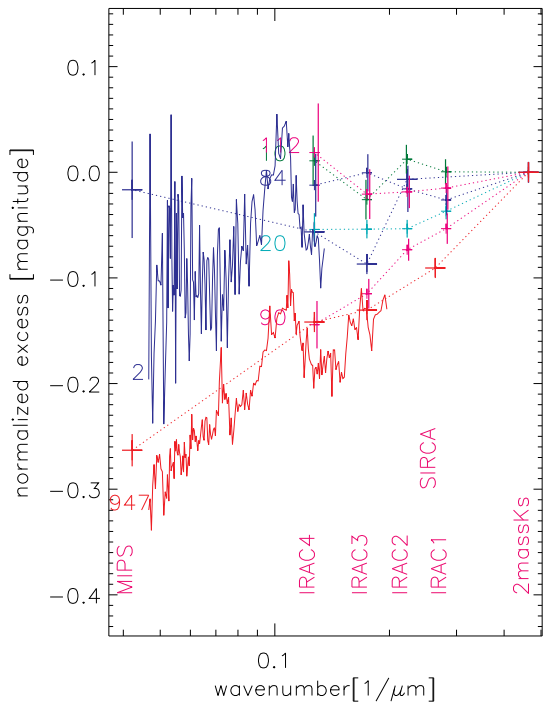}
      \caption{The "reddening" curve in the mid-IR, "normalized excess" = E($\lambda\ -\ K_s$)/E(I - K$_s$)
      vs wave number, for six stars behind the B\,335 globule.
      The normalized excess derived from the
      Spitzer/ISOCAM spectra of two of the stars,
      \#2 and \# 947, are included with the same normalization.}
      \label{excess}
   \end{figure}
%%_____________________________________________________________
%%_____________________________________________________________
%   extinction and extrapolation to zero wavenumber.
%   is4SpitzerExcess
%   displayFlag = [0,1,0]
%   funFlag     = [1,0,0,0,0]
%   SpitzerMidExcess820110122.eps                                figure 9
%-------------------------------------------------------------
  \begin{figure}
   \centering
   \includegraphics[width=9cm]{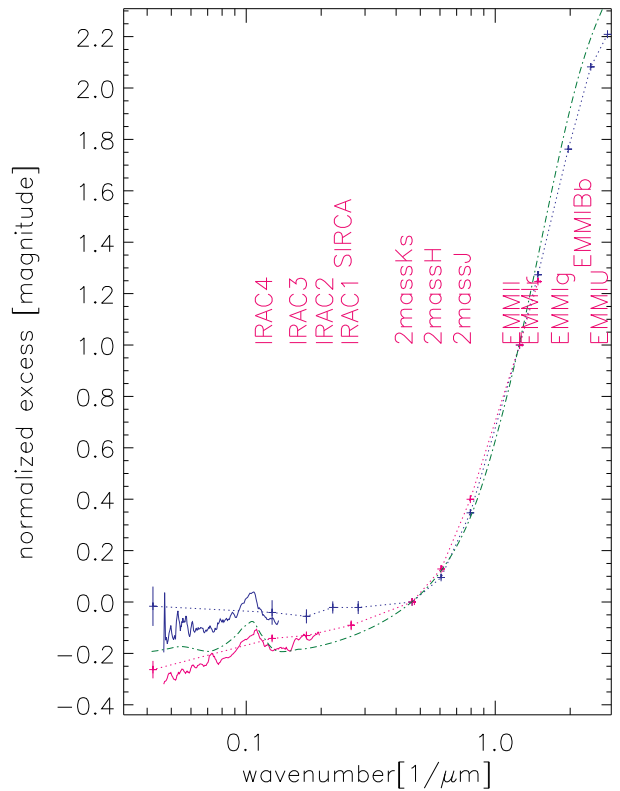}
      \caption{The normalized excess, (E($\lambda\ -\ K_s$)/E(I - K$_s$), in the mid-IR for the
      weighted average of the five stars at the rim of the B\,335 globule (blue) and the most obscured star (red).
      The green curve represents the \cite{2003astro.ph..4488D} $R_V$ = 5.5 curve.}
      \label{midExcess}
  \end{figure}
%_____________________________________________________________
%_____________________________________________________________
%%  MRNmodel0_20110218.ps                                   figure 10
%-------------------------------------------------------------
   \begin{figure}
   \centering
   \includegraphics[width=8cm]{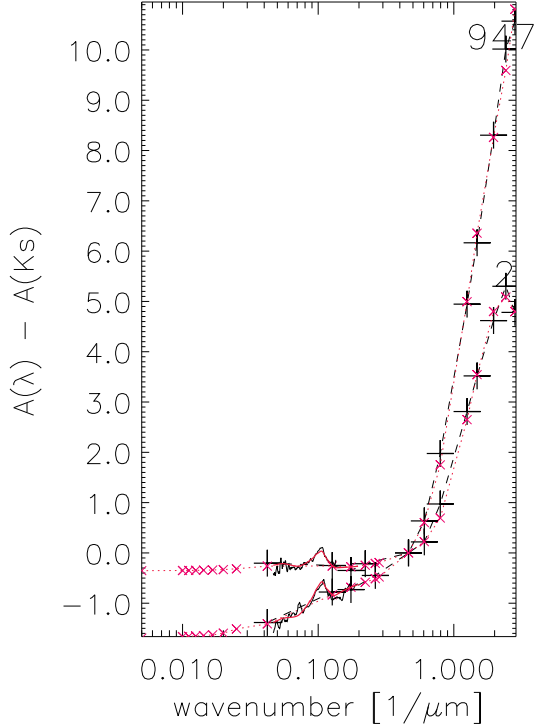}
      \caption{The excesses from the stars \#2 and \#947  and
      their corresponding model for the whole spectral range.
      Photometric measurements are marked with '+' signs,
      while the spectra are drawn as \emph{full drawn} lines.
      The models are marked with 'x' signs and \emph{red} lines.}
      \label{model0}
   \end{figure}
%%_____________________________________________________________
%_____________________________________________________________
%   extinction and extrapolation to zero wavenumber.
%   excessMin
%   funFlag = [1,0,0,0]
%   ishow = 7
%%  MRNmodel7_20110218.ps                                   figure 11
%-------------------------------------------------------------
   \begin{figure}
   \centering
   \includegraphics[width=8cm]{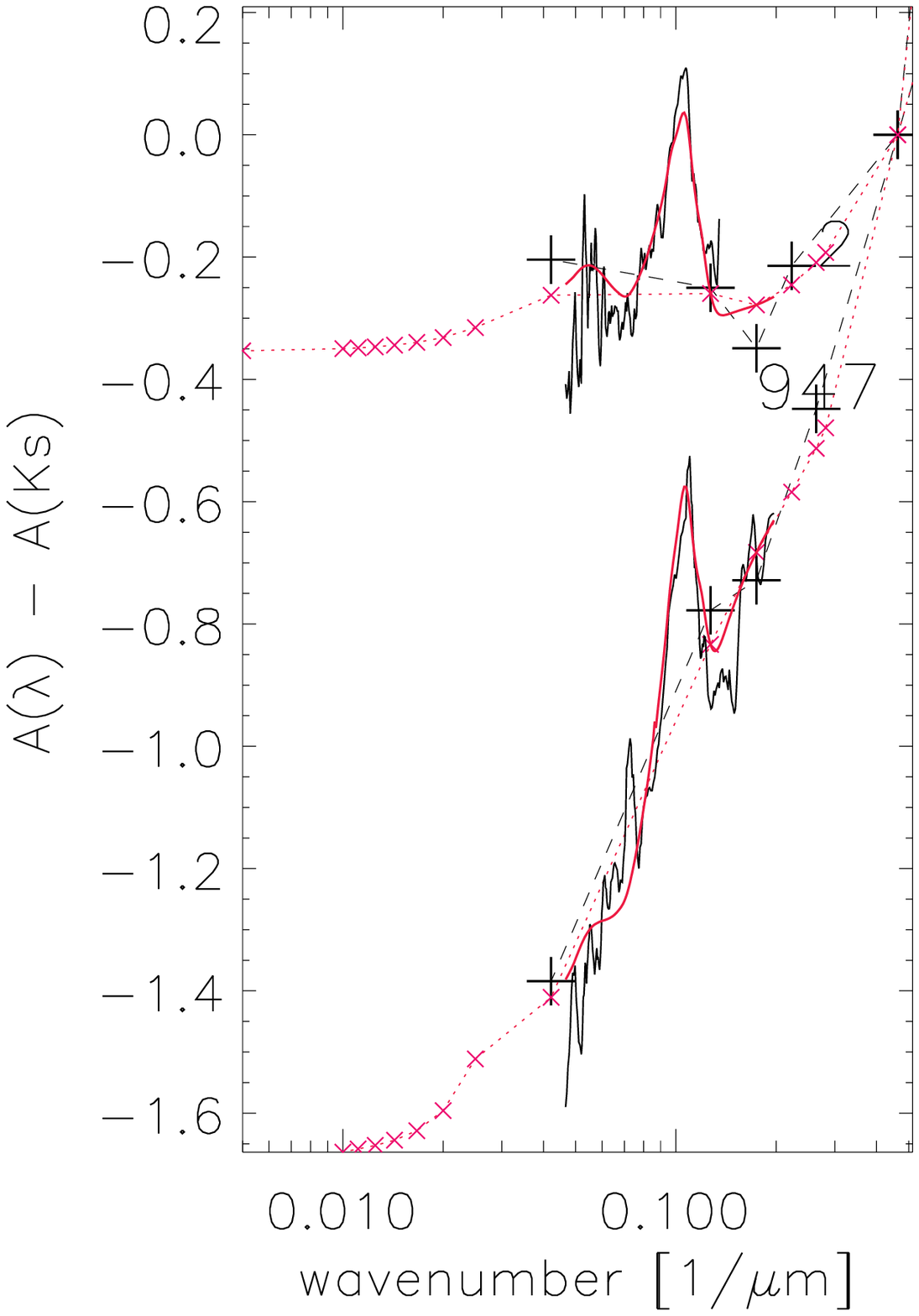}
      \caption{The excesses from the stars \#2 and \#947 and
      their corresponding model for the wavelength range $\lambda\ >\ 2\ \mu m$.
      Photometric measurements are marked with '+' signs.
      The spectra are drawn as \emph{full drawn} lines.
      The models are marked with 'x' signs and \emph{red} lines.}
      \label{model7}
   \end{figure}
%______________________________________________________________
%

%%\begin{table*}
%%\caption{Band strengths derived columndensities.}
%%\label{band}
%%\renewcommand{\footnoterule}{}  % to avoid a line before footnotes
%%\begin{tabular}{lrr}
%%\hline\\
%%
%%\textbf{measure}        &  \multicolumn{2}{\textbf{ice}}\\
%%                        &  H_{2}O  &    CO     \\
%%     W (band strength)  &  39.6    &    3.0    \\
%%     $\tau$ (peak)      &  0.305   &    0.66   \\
%%     m_{W} $[\mu g$     &  5.9     &   12.6    \\
%%\end{tabular}
%%\end{table*}

\subsection{The silicate absorption band}
The B\,335 globule was observed with the ISOCAM/CVF
and the background M giant (\#947) is the only star visible in the field.
The LW part of the scan consists of two spectral scans (back and forth)
 and we have adopted the average of the two.
The wavelength coverage is $5\,<\,\lambda \,<\ 16\,\mu$m, and except for the longest wavelengths,
the S/N is good, see Fig.\ref{fluxes}.\\
\\
Two of the background stars (\#2 and \#947) have also been observed
by Spitzer/IRS in one partial band each, Fig \ref{fluxes}.
The Spitzer/IRS and the ISOCAM/CVF spectra of star \#947 have been combined to one spectrum.
We note that the band peaks at 0.109 $\mu$m$^{-1}$ (= 9.2\,$\mu m$) with an FWHM of $\sim\ 380\ \mu m$.
The silicate feature peak is thus found at a shorter wavelengths than
e. g. described by \cite{1984ApJ...285...89D} and \cite{2004ApJS..151...35G},       %% Gibb&2004
while the FWHM-widths agree with those found for star forming regions by \cite{2004ApJS..151...35G}.
In a recent investigation, \cite{2011ApJ...731....9C},
who investigate absorption bands in the quiescent cloud IC\,5146,
also show that the silicate band is slightly shifted towards a shorter wavelength.
It is also interesting to note that even though \#947 is much more obscured than \#2,
the silicate bands have approximately the same strength.
This agrees with the results of \cite{2011ApJ...731....9C}.\\
\\
\subsection{The CO$_2$ band at 15.2 $\mu$m}
There is an indication in the CVF spectrum of the  CO$_2$ band at 15.2\,$\mu $m but it is not
confirmed in the IRS spectrum,
which has a higher S/N ratio. We can therefore only give an upper limit $\tau_{CO_{2}}\ <\ 0.15$,
corresponding to a CO$_{2}$-column density $N_{CO_{2}}\ <\ 3\cdot10^{17}\ [cm^{-2}]$.\\        %$<\ 20\ [\mu g\ cm^{-2}]$.
% \emph{Detta \"{a}r alog(medelv\"{a}rdet) - alog(medelv\"{a}rdet - 3 stddev).
% Spitzer-IRS tables give error in each wavelength of the spectra,
% to be $\tau_{CO_{2}}\ <\ 0.15$}.
% According to White&2009ApJS180_182labCO2ice.pdf the width of the CO$_{2}$ is 20 $cm^{-1}$.
%%
%%_____________________________________________________________
\subsection{The reddening curve in the mid infrared}
For each star in our sample we compare the observed colour indices to that calculated without
interstellar reddening. We refer the colour excess to the K$_s$ band
\begin{displaymath}
 E(\lambda \ -\ K_s)\ =\ (m_{\lambda} \ -\ K_s)_{observed}\ -\ (m_{\lambda} \ -\ K_s)_{intrinsic}\\
\end{displaymath}
\noindent and normalize by dividing with E(I - K$_s$).
The result is shown in Fig. \ref{excess}, where we also include the Spitzer spectra.
 The general impression is that there is not much reddening beyond the K$_s$ band,
 but in view of the uncertainties involved,
 both related to the observations and the model atmospheres, we cannot
 claim that the scatter in the diagram reflects
 real qualitative variations of the extinction in the intervening cloud.
 Even though we cannot either exclude such variations,
 it is in our view meaningful to calculate an
 error-weighted average for the "rim" stars compared to that of the most obscured stars.
 The result is shown in Fig. \ref{midExcess} and we find that the extinction is
 only going down marginally between the K$_s$ band and the silicate band.
 There is an apparent difference between the extinction in the "rim" region
 compared to the deeper part of the globule (represented by the direction towards star \#947).
 In view of the uncertainties involved, it is not quite clear that this difference is real,
 but  in the following we assume that it in fact {\it is} real and model the two cases separately.

\section{Interpretation}
\subsection{Extinction and grain properties}
\subsubsection{Grain size distribution}
\cite{1977ApJ...217..425M} (henceforth MRN) constructed grain models for                                 %% MRN 1977
the ISM extinction with two simple
grain size distributions dN(a)/da for silicates and graphite respectively,
both with dN(a)/da $ \propto$ a$^{-3.5}$ (with a as the grain radius) and $0.001\ <\ a\ < 0.25\ \mu m$.
Several attempts to find the grain size distribution of the ISM
from the measured extinction have been done.
\cite{2001ApJ...548..296W} have elaborated on this                                       %% Weingartner & Draine 2000
and suggested size distributions for graphite and silicate grains
based on more than ten parameters and explaining a large wavelength range of the ISM-extinction.
We followed their approach to analyze the B335 extinction curve
in the optical and near-IR wavelength range \citep{2010A&A...522A..84O}.\\
\\
The same approach in the wavelengths beyond 2 $\mu m$
for a molecular cloud does not seem to be that successful.
We have therefore taken a somewhat simpler approach
assuming a modified MRN grain size distribution still with spherical graphite and silicate grains and still
using the optical refractive indices as proposed by \cite{1984ApJ...285...89D} and \cite{2003ApJ...598.1017D}
(in tabulated format found at http://www.astro.princeton.edu/~draine/dust/dust.diel.html).
We also increased the grain size range to $ 0.001\ <\ a\ < 5\ \mu m$.
The distributions are assumed to follow a power law grain size distribution \emph{dN/da}
(with the spherical grain radius \emph{a}) for each substance (graphites and silicates):
\begin{displaymath}
    \frac{dN}{da}|_{MRN}  =  C_{MRN}\cdot a^{-\beta}
\end{displaymath}
without any restrictions to the parameters.
We have furthermore introduced an accumulation of grains with sizes around one grain-size
by adding a gaussian grain distribution to each of the graphite and the silicate MRN-distributions.
\begin{eqnarray*}
    \frac{dN}{da}|_{gauss} & = & 1.\ +\ C_{gauss}\ \cdot \ exp(\frac{-(a\ -\ \bar{a})^2}{a_{width}^2})\\
    \frac{dN}{da} & = &\frac{dN}{da}|_{MRN} \cdot \frac{dN}{da}|_{gauss}
\end{eqnarray*}
We use a $\chi^2$ optimisation scheme to calculate the parameter sets
$C_{MRN},\ \beta ,\ C_{gauss},\ \bar{a},\ a_{width}$ (for graphites and silicates respectively).
We find that the two different  extinction curves determined above actually can be accurately modelled
by this kind of grain size distribution (Figs. \ref{model0} and \ref{model7}).
Thus, there is no need to include any other major dust component
than silicate and graphite grains.
The corresponding size distribution of the grains are shown in  Fig. \ref{distribu}.
Even though we cannot draw too far-reaching  conclusion about the details of the size distributions,
we note that both silicate and graphite grains tend to pile up at a single grain size.
However, the peak for the graphite grains is located around 2\,$\mu$m
for the more obscured star (\#947) and at 0.5\,$\mu$m for the star closer to the rim of the globule (\#2).
There is also a striking difference in the graphite/silicate ratio, 1.6 and 0.2 respectively.
This large difference may not be real, even though it seems plausible
that the grain growth inside the globule is mainly due to gas phase carbon and PAH capture onto the graphite grains. \\
\\
With the column density of the silicates from the grain size distribution,
the solar abundance ratio [Si]/[H] (= 3.6$\cdot10^{-5}$),
and the assumption, that all silicon are found in the silicates,
the column densities N(2$\cdot H_2$ + H) can be found.
For the rim of the globule the column density ratio N(2$\cdot H_2$ + H)/A$_V\ =\ 2.3\cdot 10^{21}$
and towards the more obscured star the ratio is estimated to be $1.5\cdot 10^{21}$.
This is in accord with the average column density ratios
found for the globule rim earlier (in Paper II).\\

\subsubsection{The extinction curves}
So far we have only considered the {\it reddening curves},
but now we can use the models to determine the {\it extinction curves},
which requires  an extrapolation to zero wave number.
The two models give A(K$_s$)/E(I-K$_s$) = 0.34 for star \#947 and 0.13 for star \#2.
%In Table \ref{AKs} we give the normalized extinction towards the two stars.
The difference is striking - but is it reasonable?
It is clear that the extinction determined towards \#2 is more sensitive to systematic effects simply
because the obscuration is less and thereby small errors
due to how well the model atmosphere represents the true spectrum of the star,
how accurate the photometry is calibrated etc will be more important.
We also note that there is a small decline in the UV in the reddening curve for \#2,
which does not seem very likely (this feature is actually
causing the sharp peak in the derived graphite grain size distribution).
On the other hand, it is interesting to note
that both these reddening curves can accurately be represented
by the extinction of only graphite and silicate grains with matching size distributions.
From this point of view, there is no need for another dust component to explain
the well documented slow decline of the reddening curve beyond 2\,$\mu$m.
The dust model for the rim of the cloud is dominated by small grains,
which - if it is basically correct - means
that there is in fact {\it less} extinction at
and beyond the K band than previously thought.
Consequently, the reason why the reddening curve declines slowly beyond the K band
is simply because there is not much extinction left.  \\
\\
As indicated above, the reddening curve for the more obscured star (\#947)
is better determined and the corresponding extinction model should be trustworthy.
In this case, the explanation why the reddening curve declines slowly beyond the K band
is the presence of larger (graphite or more general, carbon dominated) grains.
This is not very surprising.

%%_____________________________________________________________
%\begin{table}
%\begin{minipage}[t]{\columnwidth}
%\caption{Relative extinction $A_\lambda /A_Ks$.}
%\label{ext}
%%\begin{tabular}{lr@{$\,\pm$}lrr}
%\begin{tabular}{lrrr}
%\hline \hline\\
%\textbf{wavelength}   & \multicolumn{3}{l}{\textbf{extinction} }\\
%$\mu m$             & \#2          & \#947 & power-law $\lambda^{-1.6}$\\
%\hline\\
%   0.354 & 14.49 & 8.10\\
%   0.413 & 14.88 & 7.49\\
%   0.509 & 14.11 & 6.60\\
%   0.673 & 10.96 & 5.16\\
%   0.798 &  8.55 & 4.20\\
%   1.258 &  3.24 & 2.05\\
%   1.650 &  1.70 & 1.38\\
%   2.157 &  1.00 & 1.00 & 1.000\\
%   3.550 &  0.58 & 0.66 & 0.451\\
%   4.493 &  0.52 & 0.57 & 0.309\\
%   5.731 &  0.52 & 0.50 & 0.209\\
%   7.872 &  0.39 & 0.40 & 0.126\\
%   23.70 &  0.52 & 0.25 & 0.022\\
%   1000. &  0.00 & 0.00 & 0.000\\
%\hline
%\end{tabular}
%\end{minipage}
%\end{table}
%%\cite{2003astro.ph..4488D}, \cite{2005ApJ...619..931I}, \cite{2007ApJ...663.1069F}, \cite{2007ApJ...664..357R}, %%\cite{2009ApJ...690..496C}, \cite{2011ApJ...729...92B}, \cite{2011A&A...527A.141C}
\subsection{Ices and band strengths}
The integrated optical depths, $\int \tau(\nu)d\nu$, of the
water and CO ice bands were calculated. These quantities are directly related
to the column densities via
\begin{displaymath}
N\ =\ \frac{\int \tau(\nu)d\nu}{A},
\end{displaymath}
where A is the band strength. \cite{1995A&A...296..810G}                    %% Gerakines 1995
have investigated the band strengths
for the stretching modes of O-H in water ice at $\lambda\,=\,3.02\,\mu m$
and $C\equiv O$ in CO ice at $\lambda\,=\,4.67\,\mu m$, and found
that they are not highly dependent on the pollution from other substances.
The resulting column densities of water and CO ice are listed in Table \ref{icemass}.
%_____________________________________________________________
%   S947 extinction with H$_{2}$Oice
%   is4PWDoptMin.pro
%   C:\is4\ps\WDoptMinH$_{2}$OXt_20091207.ps
%-------------------------------------------------------------
%   \begin{figure}
%   \centering
%   \includegraphics[width=9cm]{C:/tex/S947_201001/./bilder/WDoptMinXtH2O2_20100213.ps}
%      \caption{Extinction with H$_{2}$O ice band and grain size model fitted.}
%      \label{H$_{2}$Oicefit}
%   \end{figure}
%%_____________________________________________________________

%_____________________________________________________________

\begin{table*}
\begin{minipage}[t]{\columnwidth}
\caption{Ice column estimates.}
\label{icemass}
\renewcommand{\footnoterule}{}  % to avoid a line before footnotes
\begin{tabular}{lrrrrrr}
\hline\hline
%% Harjunp\"{a}\"{a}2004
\multicolumn{6}{l}{\emph{CO column density from submillimeter measurements by Harjunp\"{a}\"{a} et al (2004)}}\\
               &                     &                     &                   &         &       & \textbf{ice mass}\\
\textbf{isotop}& \textbf{N(isotop)$\cdot 10^{-14}$} &  \textbf{abundance$\cdot 10^2$} & \textbf{@$A_{J}$} & \textbf{CO gas} & \textbf{CO total} & \textbf{expected}\\
\cline{5-7}
         &$[cm^{-2}]$ &            &           &  \multicolumn{3}{c}{\textbf{column density} [$\mu g\, cm^{-2}$]}\\
\hline
\multicolumn{6}{l}{\emph{$A_J$ = 3.57 as for pos A12 Harjunp\"{a}\"{a} et al (2004)}}\\
$^{13}CO$ gas   &      45.0       &  1.108   &  3.57     &  19      &  \\
$^{13}CO$ total &     108.6       &  1.108   &  3.57     &          &  46  & 27\\
\multicolumn{6}{l}{\emph{$A_J$ = 3.45 as for sightline towards \#947}}\\
$^{13}CO$ gas   &      43.5       &  1.108   &  3.45     &  18      &  \\
$^{13}CO$ total &     105.0       &  1.108   &  3.45     &          &  44  & 26\\
\hline
\multicolumn{6}{l}{\emph{$A_J$ = 3.57 as for pos A12 Harjunp\"{a}\"{a} et al (2004)}}\\
$C^{18}O$ gas   &     11.0        &   0.204  &  3.57     &  25\\
$C^{18}O$ total &     14.5        &   0.204  &  3.57     &          & 33   &  8\\
\multicolumn{6}{l}{\emph{$A_J$ = 3.45 as for sightline towards \#947}}\\
$C^{18}O$ gas   &     10.6        &   0.204  &  3.45     &  24\\
$C^{18}O$ total &     14.0        &   0.204  &  3.45     &          & 32   &  8\\
\hline\hline
%% Gerakines1995 columnDensity=N_{H_{2}O}*mol_{H_{2}O}/Avogadro=72e16*18./6.022e23 [g cm^-2]
\multicolumn{6}{l}{\emph{ice column density using laboratory band strength measurements by Gerakines et al(1995)}}\\
\textbf{ice}  & \textbf{bandstrength$\cdot 10^{16}$} & $\int\tau(\nu)\,d\nu$& \textbf{N$\cdot10^{-16}$} &   &   &\textbf{ice mass}\\
              & [cm]   & [$cm^{-1}$]  &  $[cm^{-2}]$ & \multicolumn{3}{r}{\textbf{column density} [$\mu g\, cm^{-2}$]}\\
\hline
H$_{2}$O ice  &   2.00                & 100                  &  61        &    &   & 15\\
CO ice        &   0.11                &  3.4                 &  30        &    &   & 15\\
CO$_{2}$ ice  &   0.11                & $<$3                 &  $<$30     &    &   &$<$20\\
%% 655 cm^{-1} White&2009ApJS180_182labCO2ice.pdf width 20.
\hline\hline
%% model
\multicolumn{6}{l}{\emph{ice coat and grain size distribution gained }}\\
\multicolumn{6}{l}{\emph{from extinction measurements fitted to a modified MRN model}}\\
\textbf{ice} &\textbf{refractive} & \textbf{ice mantle } & \textbf{volume $\cdot10^{6}$} &   &    & \textbf{ice mass}\\
            &\textbf{indices}    & [$\AA$]            & [cm] &  \multicolumn{3}{r}{\textbf{column density} [$\mu g\, cm^{-2}$]}\\
\hline
H$_{2}$O ice & Fig \ref{ice} \footnote{\cite{2008JGRD..11314220W}}
                                                         & 16                &  29  &   &    & 28\\        %% Warren&Brandt2008
CO ice       & Fig \ref{icemix} \footnote{\cite{1997ApJ...479..818E}}
                                                         & (mixture) 18      &  21  &   &    & 23\\        %% Elsila&1997
\hline\hline\\
\textbf{silicate band}  &\multicolumn{1}{c}{\textbf{$\hat{\tau}_{Si}$}} & \multicolumn{2}{c}{\textbf{column density}} & &\textbf{$(\hat{\tau}_{Si}$/volume) $\cdot10^{-3}$}  &\textbf{band width}\\
\cline{3 - 4}
& $max(\tau_{silicate})$  & \textbf{volume $\cdot 10^6$}& \textbf{mass}        &   &                 & \textbf{$\int\tau(\nu)\,d\nu$}\\
&                         & [cm]                        &[$\mu g\, cm^{-2}$]   &   & $[cm^{-1}]$     & $[cm^{-1}]$\\
\hline
\#2                    & \multicolumn{1}{c}{0.40}  &   33 &  115   &      & 12  & 110    \\        %% Warren&Brandt2008
\#947                  & \multicolumn{1}{c}{0.45} &   44 &  150   &      & 10  & 110    \\        %% Elsila&1997
\hline\hline\\
\end{tabular}
\end{minipage}
\end{table*}
%_____________________________________________________________
%%   grainsizedistribution
%%   excessMin
%%   funFlag = [0,1,0,0]
%%  MRNdistribution_20110218.ps                               figure 12
%%-------------------------------------------------------------
   \begin{figure}
   \centering
   \includegraphics[width=8cm]{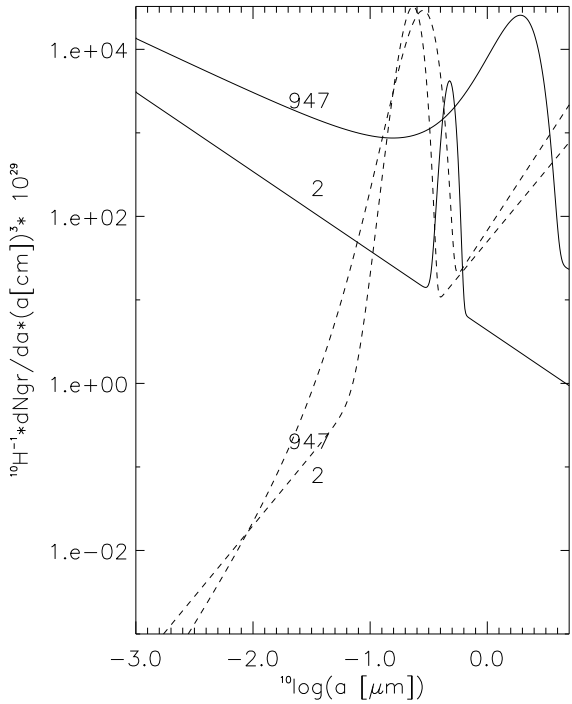}
      \caption{The grain volume distribution as modelled for the rim (probed by star \#2) and the central region of B\,335 (probed by star \#947).
      The grain size distribution model is a power-law distribution with simple
      spherical grains of graphite (\emph{full} lines) and
      silicates (\emph{dashed}) with an additional
      gaussian grain accumulation at one grain size.}
      \label{distribu}
   \end{figure}
%
%_____________________________________________________________

%   S947 grain size distribution from
%   SpitzerExcess & PWDexcessAMin
%   displayFlag = [1,0,0]
%   funFlag     = [0,0,0,1]
%   	ice			= 'H2O'
%	coatText		= 'add'
%   'H2O':aaParsExt	= 7.e-3                                   figure 13
%-------------------------------------------------------------
   \begin{figure}
   \centering
   \includegraphics[width=8cm]{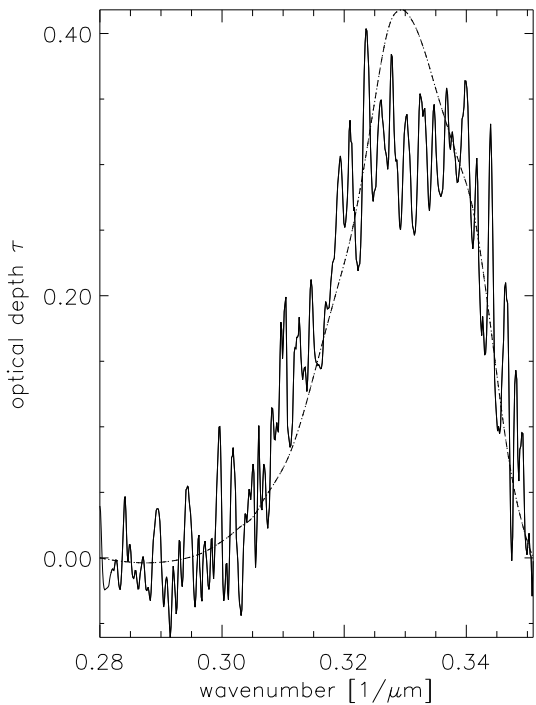}
      \caption{Detailed view of H$_{2}$O ice band (\emph{fully drawn} and a model (\emph{dashed})
      of ice coated grains \citep[with refractive indices from][]{2008JGRD..11314220W}.}       %% Warren&Brandt2008
      \label{ice}
   \end{figure}
%_____________________________________________________________
%   S947 grain size distribution from
%   is4PWDoptMix.pro    ice = 'H2O' Qtext = 'add' funFlag = 3
%   QextMix = 'TAB01'
%   C:\is4\ps\WDoptMinXtH2Oinfluence_20100310.ps           figure 14
%-------------------------------------------------------------
   \begin{figure}
   \centering
   \includegraphics[width=8cm]{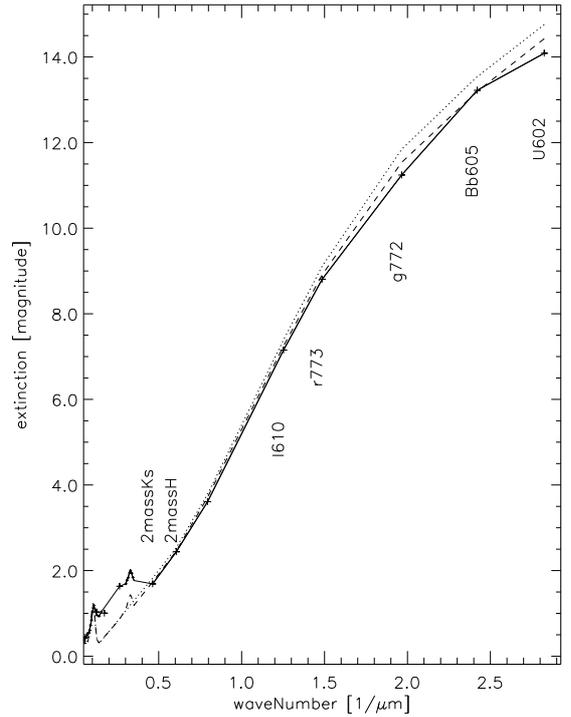}
      \caption{Influence of H$_2$O ice on the extinction in the optical wavelength range.
      The ice (\emph{dashed}) lowers the extinction in the optical
      from the ice free grain extinction (\emph{dotted}) line.}
      \label{influence}
   \end{figure}
% %_____________________________________________________________
%   \begin{figure}
%   \centering
%   \includegraphics[width=9cm]{./bilder/WDoptMinXtH2O0_20100413.ps}
%   \caption{Extinction with H$_{2}$O ice band and grain size model fitted.}
%   \label{WDopt}
%   \end{figure}
%%_____________________________________________________________
%   S947 grain size distribution from
%   is4SpitzerExcess funFlag=[0,0,0,1]
%   is4PWDexcessAMin.pro    ice = 'CO'
%   Qtext = 'add'
%   QextMix = ['TAB81','TAB85','TAB77']
%   QextMix =['TAB83','TAB89','TAB75']
%   C:\is4\ps\WDXtCOsurface0refrac_20100613.ps
%   C:\is4\ps\WDXtCOsurface0refrac_20110124.ps             figure 15
%-------------------------------------------------------------
   \begin{figure}
   \centering
   \includegraphics[width=9cm]{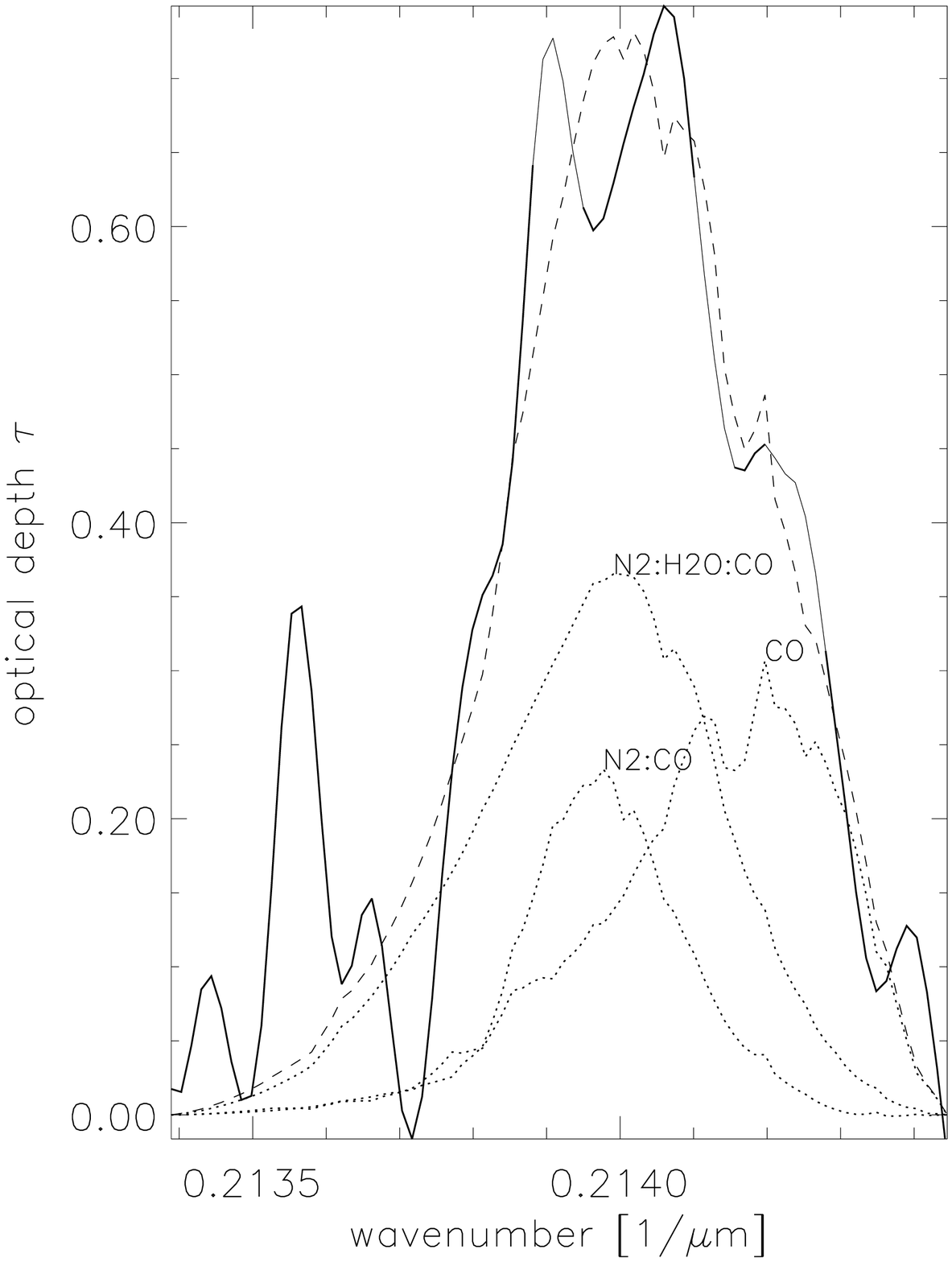}
      \caption{Detailed view of CO ice extinction and a model (\emph{dashed})
      and the different ices \citep[from][]{1997ApJ...479..818E},
      that have been included.}      %%Elsila&1997
      \label{icemix}
   \end{figure}

\subsection{CO gas column density}
\cite{2004A&A...421.1087H} have observed  $^{13}CO$ and $C^{18}O$ emission from the B\,335 globule
and in the neighbourhood of the sightline    %%% Harjunp\"{a}\"{a} 2004
to our background star (A12 in their Table 1) they found a gas N(CO)-column density,
that deviates from a linear correlation with the $A_{J}$ extinction,
probably caused by freeze-out of CO.
The gas column densities were estimated for the two isotopologues  (Table \ref{icemass}),
giving slightly different results.
%_____________________________________________________________
\begin{table}
\begin{minipage}[t]{\columnwidth}
\caption{Column density and mass ratios observed towards star \#947.}
\label{ratios}
%\begin{tabular}{lr@{$\,\pm$}lrr}
\begin{tabular}{lrrr}
\hline \hline\\
\textbf{mass ratio}        & \multicolumn{2}{c}{\textbf{column density ratios}} \\
\cline{2-4}\\
                           & \#2        & \#947 & \textbf{ISM$^a$} \\
\hline\\
graphite/silicate          & $0.2^b$    & $1.6^b$   & $ 0.4$\\
$H_{2}O_{ice}$/silicate    &            & $0.18^b$\\
$CO_{ice}$/silicate        &            & $0.15^b$\\
$CO_{ice}/H_{2}O_{ice}$    &            & $1.0^b$ \\
                           &            & $0.8^c$ \\
$N(CO_{ice})/N(CO_{gas})$  &            & 0.8 - 1\\
\hline
\end{tabular}
\footnotetext[1]{Abundance limited ratio (= 1.81) such that all C-atoms exist as graphite
and all Si-atoms are incorporated into silicates.}
\footnotetext[2]{from modelling}
\footnotetext[3]{from band strengths}
\end{minipage}
\end{table}%%
\subsection{Extinction models for ice coated grains}
The ice coatings may either form on the grains by surface reactions or
condense onto the surfaces out of the gas phase.
In both cases it is reasonable to assume that the ice thickness of the layers
is the same for all grain sizes and thus the ice mass is dependent on the size of the grain surface.
For water ice we use the refractive index from \cite{2008JGRD..11314220W} and
using the thickness of the ice layer as a parameter we find           %% Brandt&Warren2008
a good fit to the measurements as can be seen in Fig \ref{ice}.
In these calculations we have excluded grains smaller than 25\,$\AA$, for which the ice mantles evaporate  because of the instant temperature rise when hit by UV photons or cosmic rays \citep{2009ApJ...690.1497H}.

The thickness of the coating in combination with
the grain size distribution we derived above (see Fig \ref{distribu})
now allows us to calculate the H$_{2}$O column density.
We note that there is more ice needed (almost a factor 2,  see Table \ref{icemass}) to explain the absorption band when we distribute the ice on the grain surfaces than if we simply consider the band strength (which is equivalent to a thin sheet of water ice). We have not looked into the detailed reason for this difference, but it is probably real and if so,  all estimates in the literature on column densities of ices are underestimated by about a factor two!

It is interesting to note that the ice coating slightly changes extinction
also outside the absorption band and the effect is to {\it lower} the extinction, see Fig \ref{influence}.\\
\\
For the CO ice the situation is a bit more complicated because
the dielectric constants are heavily influenced by the presence of other frozen molecules,
like H$_{2}$O, O$_{2}$, N$_{2}$ as well as CO$_{2}$.
Both the peak position and the width of the CO ice band are
dependent on the mixture and in an attempt to match the observed band we have optimised
for three different mixtures using data from \cite{1997ApJ...479..818E}, see Fig \ref{icemix}.                   %% Elsila1997
We realize that this fit cannot be used to claim
that e.g. there must be N$_{2}$ frozen
onto the grains -- our fit should rather be seen as an example of a possible mixture.
However, the calculated column density is not very dependent on the exact mixture
and (like in the case of water ice) there is more CO ice needed to explain the absorption band when it is located on dust surfaces than achieved from a direct calculation using the band strength, see Table \ref{icemass}.\\
\\

\begin{table*}
\begin{minipage}[t]{\columnwidth}
\caption{Reddening (A$_\lambda$-A$_{K_s}$)/E(J-K$_s$)}
\label{reddening}
\begin{tabular}{llllllllllllll}
\hline\hline\\
\textbf{wavelength} [$\mu m$]
                     & 0.35 & 0.41 & 0.51 & 0.67 & 0.80 & 1.26 & 1.65 & 2.16 & 3.55 & 4.49 & 5.73 & 7.87 &23.7 \\
\hline\\
1 \textbf{B335 \#2}  &6.91 &7.33 &6.91 &5.11 & 3.82 & 1.00& 0.32 & 0.00  & -0.28 & -0.35& -0.40 & -0.37 & -0.38 \\
2 \textbf{B335 \#947}& 6.18 & 5.49 & 4.73 & 3.64 & 2.85 & 1.00 & 0.35 & 0.00  &-0.27 & -0.33 & -0.39 & -0.48 & -0.81 \\
\hline\\
%%\cite{2003astro.ph..4488D} \cite{2005ApJ...619..931I} \cite{2007ApJ...663.1069F} \cite{2007ApJ...664..357R} \cite{2009ApJ...690..496C}
%% \cite{2011ApJ...729...92B} \cite{2011A&A...527A.141C}
3 \textbf{\cite{2003astro.ph..4488D}}
                   &7.15 &6.67 & 5.74 & 4.09 & 3.07 & 1.00 & 0.37 & 0.00 & -0.38 & -0.49 & -0.56 & -0.55 & -0.60\\
R$_V$=5.5\\
\hline\\
4 \textbf{\cite{2005ApJ...619..931I}} & & & &    &      & 1.00 & 0.36 & 0.00 & -0.30 & -0.37 & -0.38 & -0.37\\        %%% Indebetouw 2005
\multicolumn{5}{l}{two large regions close to the Galactic plane}\\
\hline\\
5 \textbf{\cite{2007ApJ...663.1069F}} & & & &    &      & 1.00& 0.37 & 0.00 & -0.25 & -0.33 & -0.35 & -0.35 & -0.32\\ %%% Flaherty 2007
\multicolumn{5}{l}{star forming regions NGC 2068/2071}\\
\hline\\
6 \textbf{\cite{2007ApJ...664..357R}} & & & &    &      & 1.00 & 0.40 & 0.00 & -0.27 & -0.35 & -0.39 & -0.38\\        %%% Roman-Zuniga 2007
\multicolumn{5}{l}{dense cloud core}\\
\hline\\
7 \textbf{\cite{2009ApJ...690..496C}} & & & &    &      &  1.00 & 0.37     & 0.00 & -0.25 & -0.33 & -0.38 & -0.38 & -0.46\\  %%% Chapman 2009
\multicolumn{5}{l}{molecular cloud (Ophiuchus)}\\
\hline\\
8 \textbf{\cite{2011ApJ...729...92B}} & & & & &  & 1.00 & 0.35 & 0.00 & -0.26 & -0.31 & -0.35 & -0.37 & -0.44\\   %%% Boogert 2011
\multicolumn{5}{l}{69 isolated dense cores (from polynomial ibid., equation (1))}\\
\hline\\
%9 \textbf{Cambr\'{e}sy et al. (2011)} & & & &    &      &      & 1.56 & 1.0 & 0.61 & 0.50 & 0.40\\
%\multicolumn{5}{l}{Trifid nebula}\\
\hline
\end{tabular}
\end{minipage}
\end{table*}
%%

%%
%%%%%%%%%%%%
\section{Discussion}
\subsection{The reddening curve}
As mentioned, the infrared {\it reddening curve} for dark clouds
has recently been determined in a number of investigations. By {\it assuming} the extinction ratio e.g.
A$_H$/A$_K$ to be the same as for the diffuse ISM (for which it has been determined),
the {\it extinction curve} follows from:

\begin{equation}
A_\lambda /A_{K_s}\ =\ (A_H /A_{K_s}\ -\ 1)\cdot (E_{\lambda-K_s}/E_{H-K_s})\ +\ 1.
\end{equation}

As we question (see below) the assumption of a more or less universal A$_H$/A$_K$ value, we have converted the "extinction" given in these papers back to the reddening as given in Table \ref{reddening}. For comparison we also include the reddening curve for the diffuse ISM determined by  \cite{2005ApJ...619..931I}. There is a good agreement between the different investigations and the reddening of our star \#2, while our result for star \#947 differs significantly at the longest wavelengths. For reference we also give the reddening produced by the R$_V$ = 5.5 model from \cite{2003astro.ph..4488D}. As has become quite clear during the past decade, the dip in the wavelength range 3--7\,$\mu$m, predicted by this model is not present in the observations. For this reason, one cannot either trust that it can be used as a link between the near-IR reddening and the absolute extinction (which is frequently assumed).

It is interesting to note that the interstellar reddening of the diffuse ISM
does not significantly differ from that of dark clouds, at least until the IRAC\,4 band.
It is unfortunate, however, that the MIPS\,24 band has not yet been included in any investigation of the diffuse ISM.

\subsection{The extinction}
\subsubsection{The diffuse and moderately dense ISM}

In Paper\,II we studied the extinction of the same dark cloud
in the optical and near-IR regions based on a large number of stars.
We came to the conclusion that the CCM functional form represents the reddening well.
However, this does not necessarily mean that the {\it extinction curve}
is adequately described by the CCM function at these wavelengths.
The problem has to do with the extrapolation of the reddening curve to zero wave number.
Even though variations of the reddening curve in the near-IR have been noted,
it is generally found that the reddening curve can be understood
if the extinction in the near-IR follows a power law ($A_{\lambda} \propto \lambda^{-\beta}$),
where $\beta =$ 1.6 -- 1.8 \citep[see][]{2003ARA&A..41..241D}.
As cited in the Introduction there is, however, increasing evidence
that such a power law is no longer valid beyond 3\,$\mu$m,
and it is therefore {\it not} adequate to assume e.g. a A$_H$/A$_K$ ratio
based on a power law extrapolation.
The obvious solution is to extend the observations further into the infrared,
and this is the approach of the present paper.
However, even if the Spitzer archive allows us to use observations out to 24\,$\mu$m,
the extinction is not zero at this wavelength either.
Somehow we must extrapolate to zero wave number.
For this purpose we use a grain model designed
to fit the reddening curve all the way from the UV to 24\,$\mu$m,
including the important silicate absorption as well as ice features (for the most obscured star).
Our grain model is a bit {\it ad hoc} (a power law size distribution plus a Gaussian distribution
centered on a certain size kept as a parameter in the fitting process),
but it is interesting that only silicates and carbon grains
(plus ice mantles) are needed to explain the reddening curve(s)
from the UV all the way to 24\,$\mu$m, including the region 3--8\,$\mu$m.
Thus, even though we cannot claim that our grain models are unique,
they enable us to extrapolate the reddening curve to zero wave number in a physically reasonable way. \\

Ideally one would like to measure the relation between the absolute extinction
and the reddening without being dependent on a dust model.
As discussed in Paper\,II, star counts offer a way to find the true extinction at a given wavelength,
thereby providing the link between the reddening curve and the extinction curve.
The main problem in applying this method to dark clouds is their small scale structure.
In practice large uncertainties result.\\

Based on a near-IR survey of the Galactic Bulge, \cite{2006ApJ...638..839N}
identify the  Red Clump (RC) stars in the colour/magnitude diagrams (CMD) for a large number of sub-fields.
Because of variations of the extinction across the field
the position in the CMD of the peak of the RC differs from one sub-field to another,
and the locus of all these  RC peaks directly gives the A$_{Ks}$/E(J-K$_s$) ratio.
Their accuracy is very good: A$_{Ks}$/E(J-K$_s$) = 0.494 $\pm$0.006.
This is close to our result for star \#2: A$_{Ks}$ /E(J$-$K$_s$) = 0.51.
It is also interesting to note that \cite{2011ApJ...737...73F},
who determine the extinction to the galactic centre by measuring
the hydrogen recombination lines from 1.28 to 19\,$\mu$m
(relating the absolute scale to the radio continuum and using Menzel's Case B), get  A(K$_s$)/E(J-K$_s$) $\approx$A(2.16)/(A(1.28)-A(2.16)) = 0.51. \\

Three different methods thus arrive at the same A(K$_s$)/E(J-K$_s$) ratio.
All three investigations probe both the diffuse ISM and denser regions,
and one may wonder if this result is applicable for the diffuse and/or dense ISM in general.
\cite{2005ApJ...619..931I} investigate two directions
close to the Galactic plane ({\it l}= 0 42 and {\it l}= 284)
and get a different result: A(K$_s$)/E(J-K$_s$) = 0.67$\pm$0.07.
Their method is also based on  the statistical evidence
that red giant stars  have a relatively narrow distribution both regarding
their near-IR colours and their absolute magnitudes and by assuming
that the {\it diffuse} interstellar matter is relatively smoothly distributed along the lines of sight,
\cite{2005ApJ...619..931I} can interpret the colour/magnitude plots
in terms of extinction ratios and the extinction per kpc.
They base their analysis on large area surveys,
combining 2Mass with  Spitzer mapping, and the red giant population is well separated
from the dwarf population in their diagrams (see their Fig. 3). \\

\begin{figure}[t]
\centering
\includegraphics[width=9cm]{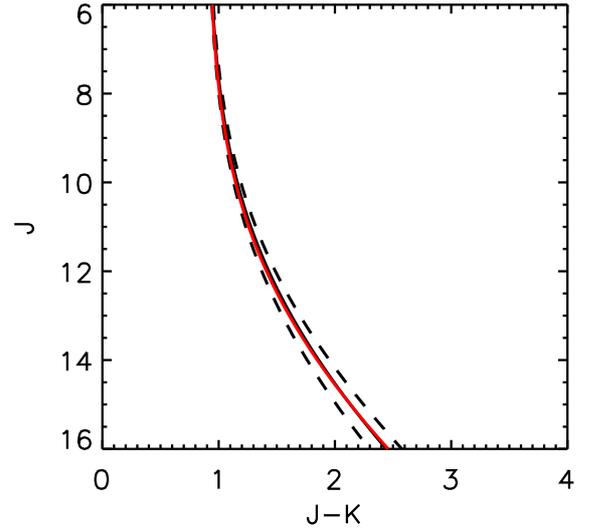}
  \caption{The black curve represents the locus of typical red giant stars
  at different distances along a line of sight with an extinction
  of c$_K$=0.15 kpc$^{-1}$ and E(J-K$_s$) = 1.5$\pm$0.2\,A$_{Ks}$ as determined
  by \cite{2005ApJ...619..931I} for the diffuse ISM.
  The dashed curves represent the error estimates by these authors.
  The red curve (on top of the full drawn black curve)
  represents our determination of star \#2 with  c$_K$=0.105 kpc$^{-1}$
  and E(J-K$_s$) = 1.96\,A$_{Ks}$.}
  \label{Indebetouw1}
\end{figure}

\begin{figure*}[t]
   \centering
   \includegraphics[width=18cm]{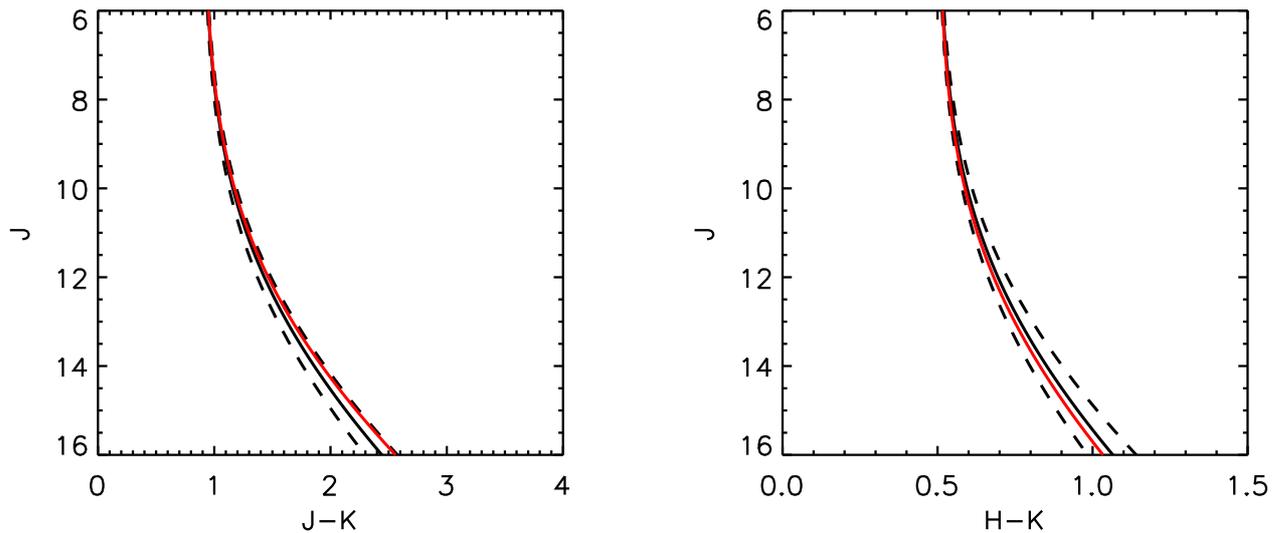}
      \caption{The same assumptions as in Fig \ref{Indebetouw2}.
      The dashed curves represent the error estimates by these authors.
      The red curves represents our determination of star \#2 with  c$_K$=0.12 kpc$^{-1}$.
      The red curves are within the error limits quoted by  \cite{2005ApJ...619..931I}.
      This is remarkable in view of the large difference in the E(J-K$_s$)/A$_{Ks}$
      and E(H-K$_s$)/A$_{Ks}$  between our result for star \#2
      (which probes the outer regions of the B\,335 globule plus diffuse clouds
      behind the globule) and the results of  \cite{2005ApJ...619..931I}.}
      \label{Indebetouw2}
\end{figure*}

Even though the extinction in our line of sight to star \#2 should be dominated by the dark cloud
there is also  a component of diffuse extinction from the path behind the star.
It is therefore of interest to investigate
whether e.g. our extinction ratio A$_J$/A$_K$=2.96 would deviate by large
from the fit to the observed red giant distribution observed by \cite{2005ApJ...619..931I}.
From their fit they get $A_J$/A$_K$ = 2.5$\pm$0.2  and c$_K$=0.15$\pm$0.1.
We use the intrinsic colour index and the absolute magnitude of a K1 giant
to construct the J versus J-K diagram for such stars along a line of sight.
By adjusting the c$_K$ value to 0.105 \cite[i.e. well within the error estimates of][]{2005ApJ...619..931I}
we can exactly reproduce the same curve, see Fig \ref{Indebetouw1}.
For the corresponding diagram with J versus H-K
where we compare the curve based on the result of  \cite{2005ApJ...619..931I}
(with $A_H$/A$_K$ = 1.5$\pm$0.1  and c$_K$=0.15$\pm$0.1) we get an identical curve for our value,
$A_H$/A$_K$ = 1.63,  if we use c$_K$=0.135.
This value differs from that we used to make the curves overlap in the previous diagram.
If we, as a compromise, use the mean value, c$_K$=0.12,
we still get curves well within the errors quoted by  \cite{2005ApJ...619..931I}, see Fig. \ref{Indebetouw2}.\\

What does this comparison tell us?
It basically illustrates the difficulty to determine the link
between the reddening and the extinction curves.
Not even such a careful investigation, based on high quality data, as that of  \cite{2005ApJ...619..931I},
excludes considerably different results for the diffuse ISM
(like our result for star \#2 or those determined for the path towards the GC by
\cite{2006ApJ...638..839N} and \cite{2011ApJ...737...73F}.

\subsubsection{Dark clouds}
The reddening curve for star \#947, sampling dense regions of the B\,335 globule, differs considerably at the longest wavelength (24$\mu$m) from previous work by
\cite{2007ApJ...663.1069F}, % Flaherty
\cite{2009ApJ...690..496C} %Chapman
and \cite{2011ApJ...729...92B}% Boogert
 (see Table \ref{reddening}). On the other hand, the results of these authors agree well to ours for star \#2, which probes a less dense region than \#947. Does it also mean that \#947 traces denser regions than those investigated in the cited papers. \cite{2007ApJ...663.1069F} analyse data from five star formation regions, and there are small differences regarding the reddening in the IRAC bands. For the MIPS 24\,$\mu$m they present data for two regions, NGC 2068/2071 and the Serpens region. Their background sources  probe regions with less extinction than that for \#947. \cite{2009ApJ...690..496C} analyse data from 3 star formation regions and they look separately at regions with different extinctions. Their most obscured stars have extinctions comparable to that of our star \#947. Finally, \cite{2011ApJ...729...92B}, who analyse a number of dense cores, certainly include lines of sight with larger extinction than our case study. They determine the spectral classes from IR spectroscopy and use stellar atmosphere models to determine the intrinsic SEDs of the stars. For one of their stars, judged to be well classified, they determine the extinction curve for the continuum assuming the same A$_{Ks}$/E(J-K$_s$) ratio as determined by Indebetouw. In Table \ref{reddening} we have removed that assumption to get the reddening curve. Consequently, the reddening curve determined by \cite{2011ApJ...729...92B} is only based on one star and their method is the same as ours. However, their reddening curve is applied to the 27 other stars and in most cases it fits well, except frequently at the longest wavelength, MIPS 24\,$\mu$m. It is also at this wavelength where our results differ significantly.

There appears to be a problem to include the MIPS 24\,$\mu$m fotometric point in a reliable way in these studies. Not only in the large sample of individually studied stars by  \cite{2011ApJ...729...92B} there are many cases which differs significantly from the others. Also in the investigation of \cite{2009ApJ...690..496C} the less obscured regions exhibit unrealistic values at 24\,$\mu$m. Assuming that the technical details (colour correction due to the broad filter, the psf correction) have been correctly done in all the cited papers, there remains three sources of error:\\

\begin{enumerate}
\item {\it The MIPS 24\,$\mu$m photometry is not reproducible enough.} According to the MIPS handbook, the repeatability is 1\% or better and the absolute calibration is better than 5\% determined (using observations of solar type stars, A, and K giant stars). This cannot be the main reason for the apparent problems with the 24\,$\mu$m photometry. \\

\item {\it The model atmospheres do not represent the stellar SED well enough for late-type giants.} This may well be an issue in particular as the model fits are done with relatively low spectral resolution, preventing an accurate match of metallicity,  microturbulence and surface gravity. The most important parameter, T$_{eff}$, is probably well determined by the combination of photometry and spectroscopy. It should be noted, however, that  \cite{2011ApJ...729...92B} do not use high resolution synthetic spectra for their comparison to the observed spectra - instead they use opacity sampled spectra and degrade the resolution of the observed spectra from R = 2000 to R = 100 for the comparison. This is in our view a waste of information and a likely source of uncertainty.\\

\item {\it Late type giants are often irregular variables (and, indeed Mira variables).}  As the observations in the various wavebands are measured at different times, this can cause artificial drops and/or rises in the observed SEDs. There is no indication in the observations of \#947 in B\,335 that this is the case - actually a smooth SED results even though seven different instruments contribute.

\end{enumerate}

\begin{figure}[t]
   \centering
   \includegraphics[width=9 cm]{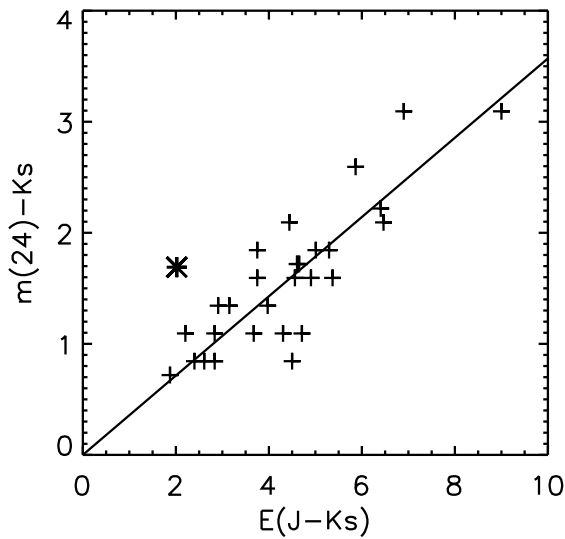}
      \caption{This colour/colour diagram is based on data from an investigation by  \cite{2011ApJ...729...92B} of field stars behind cloud cores. The slope of the curve is well defined, 0.357$\pm$0.013.  The asterisk represents the star \#947 in our investigation, and it is clearly different from the general trend.
       }
      \label{boogert}
   \end{figure}

In order to minimise the possible errors due to the stellar model spectra, we have plotted the stars in the investigation by  \cite{2011ApJ...729...92B} in a diagram with the colour index K$_s$-m(24) versus E(J-K$_s$). The E(J-K$_s$) values are taken from their Table\,3 in accordance with their assumption, E(J-K$_s$)=1.5\,A(K$_s$). In the same graph (Fig. \ref{boogert}) we also include our star \#947, and it is clear that the position of this star in the plot significantly differs from the general trend, which is actually well determined. If we neglect the (small) difference in the intrinsic colour index, (K$_s$-m(24))$_0$, for the individual stars we get\\

E(K$_s$-m(24)) = (0.357$\pm$0.013)\,E(J-K$_s$)\\

This value is lower than \cite{2011ApJ...729...92B} derived from one star
in their sample (0.44) and {\it much} lower than our value for star \#947  (0.84).\\
\\
Is there a reasonable explanation for this difference?
We first note that the estimated error due to classification uncertainty
and the photometric accuracy is within the size of the symbol of \#947 in Fig. \ref{boogert},
so it should be excluded to explain the exceptional position of this star
in the diagram by poor photometry or classification.
Even though B\,335 is not very far from the galactic plane ({\it b} = -6.5),
most of the cores investigated by \cite{2011ApJ...729...92B} are located closer
to the galactic plane and/or towards the galactic center.
This means that much of the continuum extinction can be due to the path
behind the cores and as a consequence much of the determined E(J-K$_s$)
could be due the diffuse ISM, while most of the mid-IR excess,
E(K$_s$-m(24)) could be caused by the dense cores.
Such an explanation might be plausible,
but we cannot rule out that our case study of the extinction
towards star \#947 behind a dense part of B\,335 reflects unusual cloud properties.

\section{Conclusions}
Based on photometry in the range 0.35 - 24 $\mu m$ as well as IR spectroscopy
we determine the reddening curve towards stars behind the dark globule B\,335.
For the most obscured star we also determine the absorption bands due to water ice, CO ice and silicates. -
We  confirm earlier findings that the reddening curve levels off
beyond the K band in contrast to current models \citep[e.g.][]{2001ApJ...548..296W}.
From the reddening curve we also draw the following conclusions concerning
the grain size distribution, the extinction and the ice bands.\\
\\
\texttt{Grain size distribution.}\\
\emph{The grain size distribution features.} Based on a relatively simple assumption for the grain size distribution
(a power law and a Gaussian) for silicate and graphite grains
we find solutions that closely match the observed reddening curves
for the two background stars for which we have silicate band observations.
One of the stars probe the central part of the globule and the other the rim.
For the most obscured star the model fit has a peak of graphite grains at a radius of 2\,$\mu$m,
while the graphite peak for the other star is at 0.5\,$\mu$m.
In neither case there is any need for an additional grain material to explain the reddening curve.\\
\\
\emph{The abundance ratio graphite/slicates} differs substantially for the two models (much more graphite in the central region). As a consequence we have two different explanations of the shallow reddening curve beyond the K band: for the denser region it is because of larger graphite or, more general, carbon grains which give rise to a higher mid-IR extinction than predicted by previous models constructed to match the UV to near-IR spectral range \citep[cf.][]{2001ApJ...548..296W}. For the outer region of the globule the explanation is simply that the {\it extinction as such is low beyond the K band}. Actually, much of the extinction towards the stars at the rim of the globule may originate in the diffuse medium beyond the cloud, and we suggest that this explanation to the shallow reddening curve beyond the K band is valid not only for the outer regions of the cloud but also for the diffuse ISM.\\
\\
\emph{Gas column density.} The column densities ratios $N(2\cdot H_{2}\ +\ H)/A_{V}$
are estimated to be $2.3\cdot 10^{21}$ for the lower extinction sightline
and $1.5\cdot 10^{21}$ for the higher extinction sightline.\\
\\
\texttt{Extinction model.}\\
The extinction model for \emph{the outer part} of  B\,335 gives the relation  A$_{Ks}$ = 0.51 $\cdot$E(J$-$K$_s$).
This latter relation agrees with recent determinations
towards the galactic centre by  \cite{2006ApJ...638..839N} and \cite{2011ApJ...737...73F} using different methods. \cite{2005ApJ...619..931I} get a higher value for the diffuse ISM,
A$_{Ks}$ = 0.67$\cdot$E(J$-$K$_s$), but we show that this difference is not significant
and conclude that A$_{Ks}$ $\approx$ 0.51 $\cdot$E(J$-$K$_s$) for the diffuse ISM
as well as for moderately dense molecular regions. \\
\\
The  extinction model for \emph{the denser part} of the B\,335
gives the relation A$_{Ks}$ = 0.97 $\cdot$E(J$-$K$_s$).
Although this model describes the reddening curve from the UV to 24\,$\mu$m,
including the silicate, the water-ice, and the CO-ice bands well,
we have indications that it represents conditions that may be different from other dense cores. \\
\\
\texttt{Ice bands.}\\
The \emph{water ice} band is reproduced by grains coated with pure water ice.
The column density of water ice is half of that found
in the Taurus clouds (\cite{2001ApJ...547..872W}'s) for the same extinction.                              %% Whittet et al 2001
The difference is likely due to a larger interstellar radiation field in the B\,335 region,
desorbing ice mantles deeper in the cloud.\\
\\
The \emph{CO ice} band differs in position from that of pure CO ice mantles,
indicating that CO is mixed in other ices. Based on published laboratory measurements of frozen gas mixtures
we show that the CO ice band can be understood if CO is part of such mixtures
(where H$_{2}$O is a natural part and N$_{2}$ a possible part).
The ice/gas ratio is relatively high (0.8 - 1.)
and as this is an average
along the $A_{J}\ =\ 3.5$ cloud extinction towards the probing star and
as we must allow for relatively deep regions free from CO ice on both the front
and the rear side of the cloud, the ice/gas CO ratio must be
greater than unity in the interior of the cloud, and more than half of the gas is depleted.
This confirms earlier findings \citep{2002A&A...389L...6B} that CO (or CO isotopologues) cannot be used
as a mass tracer in dense and cold molecular clouds.\\
\\
The \emph{CO$_{2}$ ice} band at 15.2 $\mu $m indicated in the ISOCAM-CVF spectrum is not confirmed
in the Spitzer spectra and we can only give
an upper limit to the CO$_{2}$-ice column density.\\
\\
%%\begin{table}
%%\caption{Column densities based on Si abundance and all Si in silicates.}
%%\label{coldens}
%%\renewcommand{\footnoterule}{}  % to avoid a line before footnotes
%%\begin{tabular}{rrrrr}
%%\hline \hline
%%\textbf{star\#}  &  \textbf{A$_V$}   & \textbf{N(2 H$_2$ + H)/A$_V$} & \textbf{A$_{K_s}$} & \textbf{N(2 H$_2$ + H)/A$_{K_s}$}\\
%%                 &                   & $\cdot 10^{-21}\ [cm^{-2}]$ &                    & $\cdot 10^{-21}\ [cm^{-2}]$\\
%%\hline\\
%%        2        &  4.8              &  2.30                       & 0.35           & 30.8\\
%%        947      &  9.6              &  1.53                       & 1.70           & 8.6\\
%%\hline \hline
%%\end{tabular}
%%\end{table}

\begin{acknowledgements}.\\
\indent This publication makes use of data products from
the Two Micron All Sky Survey, which is a joint project of the
University of Massachusetts and the Infrared Processing and Analysis
Center/California Institute of Technology, funded by the National
Aeronautics and Space Administration and the National Science
Foundation.\\
\indent This work is based [in part] on archival data obtained with the
$Spitzer$ Space Telescope, which is operated by
the Jet Propulsion Laboratory, California Institute of Technology
under a contract with NASA.\\
\end{acknowledgements}
%%  Hauschildt colours for EMMI 2mass Spitzerfilters One column figure 1
%
\bibliographystyle{aa}          % style aa.bst
\bibliography{../../Xtinct}    % your references Yourfile.bib
%
 %\bibliographystyle{amsplain}
% \bibliography{C:/tex/Xtinct}
%\begin{thebibliography}{}
%
%  \bibitem[1966]{baker} Baker, N. 1966,
%      in Stellar Evolution,
%      ed.\ R. F. Stein,\& A. G. W. Cameron
%      (Plenum, New York) 333
%
%\end{thebibliography}

\end{document}